\documentclass[journal]{IEEEtran}
\usepackage{graphics, subfigure, color, epsfig, psfrag, amsmath, amsfonts, amssymb, eucal, balance, multirow}

\newtheorem{cor}{Corollary}
\newtheorem{lem}{Lemma}
\newtheorem{prop}{Proposition}
\newtheorem{defn}{Definition}

\DeclareMathAlphabet{\eurm}{U}{eur}{m}{n}
\DeclareMathAlphabet{\mathbsf}{OT1}{cmss}{bx}{n}
\DeclareMathAlphabet{\mathssf}{OT1}{cmss}{m}{sl}
\DeclareMathAlphabet{\mathcsf}{OT1}{cmss}{sbc}{n}

\newcommand{\randomvalue}[1]{\eurm{\uppercase{#1}}}

\DeclareSymbolFont{bsfletters}{OT1}{cmss}{bx}{n}  
\DeclareSymbolFont{ssfletters}{OT1}{cmss}{m}{n}
\DeclareMathSymbol{\bsfGamma}{0}{bsfletters}{'000}
\DeclareMathSymbol{\ssfGamma}{0}{ssfletters}{'000}
\DeclareMathSymbol{\bsfDelta}{0}{bsfletters}{'001}
\DeclareMathSymbol{\ssfDelta}{0}{ssfletters}{'001}
\DeclareMathSymbol{\bsfTheta}{0}{bsfletters}{'002}
\DeclareMathSymbol{\ssfTheta}{0}{ssfletters}{'002}
\DeclareMathSymbol{\bsfLambda}{0}{bsfletters}{'003}
\DeclareMathSymbol{\ssfLambda}{0}{ssfletters}{'003}
\DeclareMathSymbol{\bsfXi}{0}{bsfletters}{'004}
\DeclareMathSymbol{\ssfXi}{0}{ssfletters}{'004}
\DeclareMathSymbol{\bsfPi}{0}{bsfletters}{'005}
\DeclareMathSymbol{\ssfPi}{0}{ssfletters}{'005}
\DeclareMathSymbol{\bsfSigma}{0}{bsfletters}{'006}
\DeclareMathSymbol{\ssfSigma}{0}{ssfletters}{'006}
\DeclareMathSymbol{\bsfUpsilon}{0}{bsfletters}{'007}
\DeclareMathSymbol{\ssfUpsilon}{0}{ssfletters}{'007}
\DeclareMathSymbol{\bsfPhi}{0}{bsfletters}{'010}
\DeclareMathSymbol{\ssfPhi}{0}{ssfletters}{'010}
\DeclareMathSymbol{\bsfPsi}{0}{bsfletters}{'011}
\DeclareMathSymbol{\ssfPsi}{0}{ssfletters}{'011}
\DeclareMathSymbol{\bsfOmega}{0}{bsfletters}{'012}
\DeclareMathSymbol{\ssfOmega}{0}{ssfletters}{'012}

\newcommand{\rvs}{{\randomvalue{s}}}	

\newcommand{\rvu}{{\randomvalue{u}}}	

\newcommand{\rvv}{{\randomvalue{v}}}	

\newcommand{\rvw}{{\randomvalue{w}}}	

\newcommand{\rvx}{{\randomvalue{x}}}	

\newcommand{\rvy}{{\randomvalue{y}}}	

\newcommand{\rvz}{{\randomvalue{z}}}	

\begin{document}

\title{A Regression Approach to Certain Information Transmission Problems}
\author{Wenyi Zhang, Yizhu Wang, Cong Shen, and Ning Liang
\thanks{W. Zhang, Y. Wang, and N. Liang are with CAS Key Laboratory of Wireless-Optical Communications, University of Science and Technology of China, China. C. Shen is with Department of Electrical and Computer Engineering, University of Virginia, USA. Email: {\tt wenyizha@ustc.edu.cn}. The work was supported by National Natural Science Foundation of China through Grant 61722114. Some preliminary results in Sections \ref{sec:memoryless} and \ref{subsec:channel-memory} have been presented in part at Information Theory and Applications (ITA) Workshop, La Jolla, CA, USA, Jan.-Feb. 2016 \cite{zhang16:ita}.}
\thanks{\copyright 20XX IEEE. Personal use of this material is permitted.  Permission from IEEE must be obtained for all other uses, in any current or future media, including reprinting/republishing this material for advertising or promotional purposes, creating new collective works, for resale or redistribution to servers or lists, or reuse of any copyrighted component of this work in other works.}
}

\maketitle

\begin{abstract}
A general information transmission model, under independent and identically distributed Gaussian codebook and nearest neighbor decoding rule with processed channel output, is investigated using the performance metric of generalized mutual information. When the encoder and the decoder know the statistical channel model, it is found that the optimal channel output processing function is the conditional expectation operator, thus hinting a potential role of regression, a classical topic in machine learning, for this model. Without utilizing the statistical channel model, a problem formulation inspired by machine learning principles is established, with suitable performance metrics introduced. A data-driven inference algorithm is proposed to solve the problem, and the effectiveness of the algorithm is validated via numerical experiments. Extensions to more general information transmission models are also discussed.
\end{abstract}

\section{Introduction}
\label{sec:intro}

When designing information transmission systems, the commonly adopted approach has been model-driven, assuming a known statistical channel model, namely the channel input-output conditional probability law. But in certain application scenarios, the underlying physical mechanism of channel is not sufficiently understood by us to build a dependable channel model, or is known but yet too complicated to prefer a model-driven design, e.g., with strongly nonlinear and high-dimensional input-output relationship. Such scenarios motivate us to raise the question: ``How to learn to transmit information over a channel without using its statistical model?''

With a sufficient amount of channel input-output samples as training data, ideally one would expect that although not using the actual statistical channel model, the encoder and the decoder can eventually adjust their operation so as to reliably transmit messages at a rate close to the channel capacity. But this is by no means a trivial task. First, nonparametric estimation of mutual information with training data only (see, e.g., \cite{paninski03:nc}) and maximization of mutual information with respect to input distribution are both notoriously difficult problems in general. Second, the decoder needs to compute his decoding metric according to the actual statistical channel model, since otherwise the decoding rule would be ``mismatched'' and the channel input-output mutual information would not be achievable.

We remark that, another line of information-theoretic works beyond the scope of this paper considers type-based universal decoding algorithms such as the maximum empirical mutual information (MMI) decoder (see, e.g., \cite{csiszar:book} \cite[Sec. IV-B-4), p. 2168]{lapidoth98:it}), which can achieve the capacity and even the error exponent of a channel without utilizing its statistical model. But such universal decoding algorithms are less amenable to practical implementation, compared with decoding with a fixed metric, which will be considered in this paper.

In this paper, we adopt a less sophisticated approach with a more modest goal. We separate the learning phase and the information transmission phase. In the learning phase, given a number of channel input-output samples as training data, the encoder and the decoder learn about some key characteristics about the statistical channel model; in the information transmission phase, the encoder and the decoder run a prescribed coding scheme, with code rate and key parameters in the decoding rule already learnt from training data during the learning phase. Therefore, our goal is not the channel capacity, but a rate under the specific coding scheme, for which the information-theoretic topic of mismatched decoding \cite{lapidoth98:it}, in particular the so-called generalized mutual information (GMI), provides a convenient tool. We note that here the terminology of training is taken from machine learning, and is different from ``channel training'' in conventional wireless communications systems, where the statistical channel model is already known (typically a linear Gaussian channel with random channel fading coefficients) and the purpose of training is to estimate the channel fading coefficients based on pilot symbols.

Section \ref{sec:memoryless} considers a general memoryless channel with known statistical model, under Gaussian codebook ensemble and nearest neighbor decoding. We allow the channel output to be processed by a memoryless function, before feeding it to the decoder. We show that in terms of GMI, the optimal channel output processing function is the minimum mean squared error (MMSE) estimator of the channel input upon observing the channel output, namely, the conditional expectation operator. This fact motivates the consideration that with training data only, the channel output processing function should also be in some sense ``close'' to the conditional expectation operator, and establishes a close connection with the classical topic of regression in machine learning.

Section \ref{sec:learning} hence follows the above thoughts to formulate a learning problem. Unlike in regression problems where performance is measured in terms of generalization error, for our information transmission problem we are interested in code rate, which is chosen based upon training data only. In Section \ref{sec:memoryless}, we propose two performance metrics: over-estimation probability, which quantifies the probability that the chosen code rate, as a random variable, exceeds the GMI; and receding level, which quantifies the average relative gap between the chosen code rate and the optimal GMI. We develop an algorithm called $\mathsf{LFIT}$ (Learn For Information Transmission) to accomplish the aforementioned learning task, which is further assessed using numerical experiments.

Section \ref{sec:extensions} discusses several potential extensions of the basic channel model in Section \ref{sec:memoryless}, including channels with memory, channels with general inputs and general decoding metrics, and channels with state. Section \ref{sec:conclusion} concludes this paper. In order to illustrate the analytical results in Section \ref{sec:memoryless}, Appendix presents case studies for several channels, some of which exhibit strong nonlinearity. These channels are also used in Section \ref{sec:learning} for assessing the $\mathsf{LFIT}$ algorithm.

Throughout this paper, all rate expressions are in nats per channel use, and logarithms are to base $e$. In numerical experiments, rates are converted into bits by the conversion rule of $1$ nat $\approx$ $1.44$ bits.

Recently, a heightened research interest has been seen in applying machine learning techniques (notably deep neural networks and variants) in physical-layer communications, including end-to-end system design \cite{28}-\cite{kim18:nips}, channel decoding \cite{29}-\cite{liang}, equalization \cite{14}-\cite{11}, symbol detection \cite{3}-\cite{26}, channel estimation and sensing \cite{4}-\cite{23}, molecular communications \cite{18}, and so on. Researchers have accumulated considerable experience about designing machine learning enabled communication systems. Instead of exploring the performance of specific machine learning techniques, our main interest in this paper is a general problem formulation for integrating basic ingredients of machine learning into information transmission models, so that within the problem formulation different machine learning techniques can be applied and compared.

\section{Channels with Known Statistical Model}
\label{sec:memoryless}

In this section, we investigate a framework for information transmission over a memoryless channel with a real scalar input and a general (e.g., vector) output. We will discuss several potential extensions in Section \ref{sec:extensions}. The central assumption throughout this section is that the statistical model of the channel, namely, its input-output conditional probability law, is known to the encoder and the decoder. The results developed in this section will shed key insights into our study of learning based information transmission problems in Section \ref{sec:learning}, where this assumption is abandoned.

\subsection{Review of Generalized Mutual Information and An Achievable Rate Formula}
\label{subsec:review}

It is well known in information theory that, given a memoryless channel with input-output conditional probability law $p(y|x)$, $x \in \mathcal{X}$, $y \in \mathcal{Y}$, when the encoder uses a codebook where each symbol in each codeword is independently generated according to certain probability distribution $p(x)$, and the decoder employs a maximum-likelihood decoding rule, the mutual information $I(\rvx; \rvy)$ is an achievable rate, and by optimizing $p(x)$ we achieve the channel capacity $C = \max_{p(x)} I(\rvx; \rvy)$ \cite{cover06:book}. When the decoder employs a decoding rule which is no longer maximum-likelihood but mismatched to the channel conditional probability law, however, mutual information fails to characterize the achievable rate, and the corresponding analysis of mismatched decoding is highly challenging; see \cite{lapidoth98:it} \cite{scarlett14:phd} and references therein for a thorough overview. In fact, the ultimate performance limit of mismatched decoding called the mismatched capacity still remains an open problem to date, and we need to resort to its several known lower bounds; see, e.g., \cite{hui83:phd}-\cite{somekh15:it}.

In this work, our main interest is not about exploring the fundamental limit of mismatched decoding, but rather about using it as a convenient tool for characterizing the achievable rate of a given information transmission model. Following the techniques in \cite{lapidoth02:it}, in \cite{zhang12:tcom}, a particular lower bound of the mismatched capacity has been derived when $\mathcal{X} = \mathbb{R}$, under the following conditions:

(1) Under average power constraint $P$, codeword length $N$, and code rate $R$ (nats/channel use), the encoder uses an independent and identically distributed (i.i.d.) Gaussian codebook, which is a set of $M = \lceil e^{NR}\rceil$ mutually independent $N$-dimensional random vectors, each of which is $\mathcal{N}(\mathbf{0}, P \mathbf{I}_{N})$-distributed. The ensemble of i.i.d. Gaussian codebooks is called the i.i.d. Gaussian ensemble.

(2) Given a length-$N$ channel output block $\underline{y} = [y_1, \ldots, y_N]$, the decoder employs a nearest neighbor decoding rule with a prescribed processing function $g: \mathcal{Y} \mapsto \mathbb{R}$ and a scaling parameter $a$ to decide the transmitted message as
\begin{eqnarray}
\hat{m} = \mathrm{argmin}_{m \in \{1, 2, \ldots, M\}} D(m),\nonumber\\
\mbox{where}\;D(m) = \sum_{n = 1}^N \left[g(y_n) - a x_n(m) \right]^2,\label{eqn:nn-metric}
\end{eqnarray}
and $\underline{x}(m) = [x_1(m), \ldots, x_N(m)]$ is the codeword corresponding to message $m$. Note that the output alphabet $\mathcal{Y}$ is arbitrary, for example, multi-dimensional, like in a multi-antenna system. Geometrically, the right hand side of (\ref{eqn:nn-metric}) is the squared Euclidean distance between a scaled codeword point and the processed received signal point, in the $N$-dimensional Euclidean space.

The lower bound derived in \cite{zhang12:tcom} is called the GMI under the codebook ensemble in condition (1) and the decoding rule in condition (2), given by the following result.

\begin{prop}\label{prop:generic-g}
For an information transmission model under conditions (1) and (2), the information rate
\begin{eqnarray}
\label{eqn:gmi-generic}
I_{\mathrm{GMI}, g} = \frac{1}{2} \log \frac{1}{1 - \Delta_g}, \quad \mbox{where}\; \Delta_g = \frac{\left(\mathbf{E}[\rvx g(\rvy)]\right)^2}{P \mathbf{E}[g(\rvy)^2]},
\end{eqnarray}
is achievable, when the scaling parameter $a$ is set as
\begin{eqnarray}
\label{eqn:a-optimal}
a = \frac{\mathbf{E}[\rvx g(\rvy)]}{P}.
\end{eqnarray}
\end{prop}
{\it Proof:} This proposition has been stated in a slightly restricted form as \cite[Prop. 1]{zhang12:tcom}. Here we briefly outline its proof, for completeness of exposition, and for easing the understanding of some technical development in Section \ref{sec:learning}.

Due to the symmetry in the codebook design in condition (1), when considering the average probability of decoding error averaged over both the message set and the codebook ensemble, it suffices to assume that the message $m = 1$ is selected for transmission, without loss of generality. Therefore the decoding metric $D(m)$ for $m = 1$ behaves like
\begin{eqnarray}
\lim_{N \rightarrow \infty} D(1) = \mathbf{E}\left[\left(g(\rvy) - a \rvx \right)^2\right], \quad \mbox{almost surely (a.s.)},
\end{eqnarray}
due to the strong law of large numbers, where the expectation is with respect to $p(x, y)$. The GMI is the exponent of the probability that an incorrect codeword $\underline{\rvx}(m)$, $m \neq 1$, incurs a decoding metric $D(m)$ no larger than $D(1)$, and hence is an achievable rate, due to a standard union bounding argument \cite[Prop. 1]{zhang12:tcom} \cite{lapidoth02:it}:
\begin{eqnarray}
I_{\mathrm{GMI}, g} = \sup_{\theta < 0, a} \left\{\theta \mathbf{E}\left[\left(g(\rvy) - a \rvx \right)^2\right] - \Lambda(\theta) \right\}, \\
\begin{split}
\Lambda(\theta) &= \lim_{N \rightarrow \infty} \frac{1}{N} \Lambda_N(N\theta), \\
\quad \Lambda_N(N\theta) &= \log \mathbf{E}\left[e^{N\theta D(m)}\big | \underline{\rvy} \right], \forall m \neq 1.
\end{split}
\end{eqnarray}
The calculation of $\Lambda(\theta)$ is facilitated by the non-central chi-squared distribution of $(g(\rvy) - a \rvx)^2$ conditioned upon $\rvy$, following \cite[App. A]{zhang12:tcom} (see also \cite[Thm. 3.0.1]{lapidoth02:it}). We can express $I_{\mathrm{GMI}, g}$ as
\begin{eqnarray}
\begin{split}
I_{\mathrm{GMI}, g} = \max_{\theta < 0, a} \bigg\{&\theta \mathbf{E}\left[\left(g(\rvy) - a \rvx \right)^2\right] - \frac{\theta \mathbf{E}[g(\rvy)^2]}{1 - 2 \theta a^2 P}  \\
&+ \frac{1}{2} \log(1 - 2\theta a^2 P)\bigg\}.\label{eqn:gmi-maximizeJ}
\end{split}
\end{eqnarray}
Solving the maximization problem (\ref{eqn:gmi-maximizeJ}) as in \cite[App. A]{zhang12:tcom},\footnote{There is a minor error in passing from (78) to (80) in \cite[App. A]{zhang12:tcom}, but it can be easily rectified and does not affect the result.} we arrive at (\ref{eqn:gmi-generic}). The corresponding optimal $a$ is given by (\ref{eqn:a-optimal}), and the optimal $\theta$ is $-(P/2)\Big /\left[P \mathbf{E}[g(\rvy)^2] - \left(\mathbf{E}[\rvx g(\rvy)]\right)^2\right]$.
{\bf Q.E.D.}

As mentioned earlier in this subsection, there are several known lower bounds of the mismatched capacity, many of which actually outperform GMI in general. We employ the GMI under conditions (1) and (2) as the performance metric in our subsequent study, because first, its expression given in Proposition \ref{prop:generic-g} is particularly neat; second, the combination of i.i.d. Gaussian codebook ensemble and nearest neighbor decoding rule provides a reasonable abstraction of many existing coding schemes (see, e.g., \cite{lapidoth02:it} \cite{zhang12:tcom} \cite{bjornson14:it}) and is in fact capacity-achieving for linear Gaussian channels (see, e.g., the appendix); and finally, it also has a rigorous information-theoretic justification. In fact, the GMI $I_{\mathrm{GMI}, g}$ is the maximally achievable information rate such that the average probability of decoding error asymptotically vanishes as the codeword length $N$ grows without bound, under the i.i.d. Gaussian codebook ensemble in condition (1) and the nearest neighbor decoding rule in condition (2); see, e.g., \cite[pp. 1121-1122]{lapidoth02:it}, for a discussion.

\subsection{Linear Output Processing}
\label{subsec:linear-processing}

In this subsection, we restrict the processing function $g$ to be linear; that is, $g(y) = \beta^T y$ where $\beta$ is a column vector which combines the components of $y$. We denote the dimension of $\beta$ and $y$ by $p$. Noting that the GMI $I_{\mathrm{GMI}, g}$ is increasing with $\Delta_g$, we aim at choosing the optimal $\beta$ so as to maximize $\Delta_g$. For this, we have the following result.

\begin{prop}
\label{prop:linear}
Suppose that $\mathbf{E}[\rvy \rvy^T]$ is invertible. The optimal linear output processing function is the linear MMSE estimator of $\rvx$ upon observing $\rvy$, given by
\begin{eqnarray}
\label{eqn:LMMSE-g}
g(y) = \mathbf{E}[\rvx \rvy]^T \mathbf{E}[\rvy \rvy^T]^{-1} y.
\end{eqnarray}
The resulting maximized $\Delta_g$ and $I_{\mathrm{GMI}, g}$ are
\begin{eqnarray}
\label{eqn:Delta-g-linear}
\Delta_\mathrm{LMMSE} = \frac{\mathbf{E}[\rvx \rvy]^T \mathbf{E}[\rvy \rvy^T]^{-1} \mathbf{E}[\rvx \rvy]}{P},
\end{eqnarray}
and
\begin{eqnarray}
\label{eqn:GMI-linear}
I_{\mathrm{GMI}, \mathrm{LMMSE}} = \frac{1}{2} \log \frac{P}{P - \mathbf{E}[\rvx \rvy]^T \mathbf{E}[\rvy \rvy^T]^{-1} \mathbf{E}[\rvx \rvy]},
\end{eqnarray}
respectively. The corresponding scaling parameter $a$ is exactly $\Delta_\mathrm{LMMSE}$ in (\ref{eqn:Delta-g-linear}).
\end{prop}
{\it Proof:} With a linear output processing function $g(y) = \beta^T y$, we rewrite $\Delta_g$ in (\ref{eqn:gmi-generic}) as
\begin{equation}
\Delta_g = \frac{\left(\mathbf{E}[\rvx g(\rvy)]\right)^2}{P \mathbf{E}[g(\rvy)^2]} = \frac{\left(\mathbf{E}[\rvx \beta^T \rvy]\right)^2}{P \mathbf{E}[\beta^T \rvy \rvy^T \beta]} = \frac{\beta^T \mathbf{E}[\rvx \rvy] \mathbf{E}[\rvx \rvy]^T \beta}{P \beta^T \mathbf{E}[\rvy \rvy^T] \beta},
\end{equation}
noting that $\mathbf{E}[\rvx \beta^T \rvy] = \beta^T \mathbf{E}[\rvx \rvy]$ with a scalar $\rvx$. This is a generalized Rayleigh quotient \cite{horn:book}. With a transformation $\tilde{\beta} = \mathbf{E}[\rvy \rvy^T]^{1/2} \beta$, we have
\begin{eqnarray}
\Delta_g = \frac{\tilde{\beta}^T \mathbf{E}[\rvy \rvy^T]^{-1/2} \mathbf{E}[\rvx \rvy] \mathbf{E}[\rvx \rvy]^T \mathbf{E}[\rvy \rvy^T]^{-1/2}\tilde{\beta}}{P \tilde{\beta}^T \tilde{\beta}}.
\end{eqnarray}
When $\tilde{\beta}$ is the eigenvector of the largest eigenvalue of $\mathbf{E}[\rvy \rvy^T]^{-1/2} \mathbf{E}[\rvx \rvy] \mathbf{E}[\rvx \rvy]^T \mathbf{E}[\rvy \rvy^T]^{-1/2}$, $\Delta_g$ is maximized as this largest eigenvalue divided by $P$. Noting that this matrix has rank one, its largest eigenvalue is simply $\mathbf{E}[\rvx \rvy]^T \mathbf{E}[\rvy \rvy^T]^{-1} \mathbf{E}[\rvx \rvy]$, and is achieved with $\tilde{\beta} = \mathbf{E}[\rvy \rvy^T]^{-1/2} \mathbf{E}[\rvx \rvy]$, i.e., $\beta = \mathbf{E}[\rvy \rvy^T]^{-1} \mathbf{E}[\rvx \rvy]$. The results in Proposition \ref{prop:linear} then directly follow.
{\bf Q.E.D.}

Note that the denominator in the logarithm of $I_{\mathrm{GMI}, \mathrm{LMMSE}}$ in (\ref{eqn:GMI-linear}) is exactly the mean squared error of the linear MMSE estimator (\ref{eqn:LMMSE-g}) (see, e.g., \cite{poor94:book}), which may be conveniently denoted by $\mathsf{lmmse}$. So we have
\begin{eqnarray}\label{eqn:gmi-lmmse}
I_{\mathrm{GMI}, \mathrm{LMMSE}} = \frac{1}{2} \log \frac{P}{\mathsf{lmmse}}.
\end{eqnarray}

In our prior works, we have examined several special cases of Proposition \ref{prop:linear}, including scalar Gaussian channels with one-bit output quantization and super-Nyquist sampling \cite[Sec. VI]{zhang12:tcom}, fading Gaussian channels with multiple receive antennas and output quantization \cite{liang16:jsac} \cite{li17:vtc}. Here Proposition \ref{prop:linear} serves as a general principle.

For the special case of scalar output, there is an interesting connection between Proposition \ref{prop:linear} and the so-called Bussgang's decomposition approach to channels with nonlinearity. Bussgang's decomposition has its idea originated from Bussgang's theorem \cite{bussgang52:rle}, which is a special case of Price's theorem \cite{rowe82:bstj}, for the cross-correlation function between a continuous-time stationary Gaussian input process and its induced output process passing a memoryless nonlinear device, and has been extensively applied to discrete-time communication channels as well (e.g., \cite{ochiai02:tcom} \cite{orhan15:ita} \cite{bjornson14:it}). For our channel model Bussgang's decomposition linearizes the channel output $\rvy$ as
\begin{eqnarray}
\rvy = \frac{\mathbf{E}[\rvx \rvy]}{P} \rvx + \rvw,\label{eqn:bussgang}
\end{eqnarray}
where the residual $\rvw$ is uncorrelated with $\rvx$. So if we formally calculate the ``signal-to-noise ratio'' of (\ref{eqn:bussgang}), we can verify that
\begin{eqnarray}
\begin{split}
\mathsf{snr}_{\mathrm{Bussgang}} = \left(\frac{\mathbf{E}[\rvx \rvy]}{P}\right)^2 \frac{P}{\mathbf{E}[\rvw^2]} &= \frac{\left(\mathbf{E}[\rvx \rvy]\right)^2}{P \mathbf{E}[\rvy^2] - \left(\mathbf{E}[\rvx \rvy]\right)^2}  \\
&= \frac{\Delta_\mathrm{LMMSE}}{1 - \Delta_\mathrm{LMMSE}},
\end{split}
\end{eqnarray}
where $\Delta_\mathrm{LMMSE}$ is exactly (\ref{eqn:Delta-g-linear}) specialized to scalar output. Hence we have
\begin{eqnarray}
I_{\mathrm{GMI}, \mathrm{LMMSE}} = \frac{1}{2} \log(1 + \mathsf{snr}_{\mathrm{Bussgang}});
\end{eqnarray}
that is, the GMI result in Proposition \ref{prop:linear} provides a rigorous information-theoretic interpretation of Bussgang's decomposition.

\subsection{Optimal Output Processing}
\label{subsec:optimal-processing}

What is the optimal output processing function without any restriction? Interestingly, this problem is intimately related to a quantity called the correlation ratio which has been studied in a series of classical works by Pearson, Kolmogorov, and R\'{e}nyi; see, e.g., \cite{renyi59:amash}. The definition of the correlation ratio is as follows.
\begin{defn}
\label{defn:renyi-correlation-coefficient}
\cite[Eqn. (1.7)]{renyi59:amash} For two random variables $\rvu$ and $\rvv$ where $\rvu$ is real scalar valued and $\rvv$ can be arbitrary, define the correlation ratio $\Theta_\rvv(\rvu)$ of $\rvu$ on $\rvv$ as
\begin{eqnarray}
\Theta_\rvv(\rvu) = \sqrt{\frac{\mathrm{var} \mathbf{E}[\rvu|\rvv]}{\mathrm{var} \rvu}},
\end{eqnarray}
if the variance of $\rvu$ exists and is strictly positive.
\end{defn}

The following result is key to our development.
\begin{lem}
\label{lem:renyi-correlation-coefficient}
\cite[Thm. 1]{renyi59:amash} An alternative characterization of $\Theta_\rvv(\rvu)$ is
\begin{eqnarray}
\label{eqn:correlation-coefficient-extremal}
\Theta_\rvv(\rvu) = \sup_g \left|\frac{\mathbf{E}[\rvu g(\rvv)] - \mathbf{E}[\rvu] \mathbf{E}[g(\rvv)]}{\sqrt{\mathrm{var}\rvu \cdot \mathrm{var}g(\rvv)}}\right|,
\end{eqnarray}
where $g$ runs over all Borel-measurable real functions such that the mean and the variance of $g(\rvv)$ exist. The optimal $g$ which solves the maximization of (\ref{eqn:correlation-coefficient-extremal}) is given by $g(v) = c \mathbf{E}[\rvu|v] + b$ where $c \neq 0$ and $b$ are arbitrary constants.
\end{lem}
{\it Proof:} The result is essentially a consequence of the Cauchy-Schwartz inequality, and has been given in \cite[Thm. 1]{renyi59:amash}. {\bf Q.E.D.}

We can show that $\Theta_\rvv(\rvu)$ lies between zero and one, taking value one if and only if $\rvu$ is a Borel-measurable function of $\rvv$, and taking value zero if (but not only if) $\rvu$ and $\rvv$ are independent.

Applying Lemma \ref{lem:renyi-correlation-coefficient} to our information transmission model, we have the following result.
\begin{prop}
\label{prop:nonlinear}
The optimal output processing function is the MMSE estimator of $\rvx$ upon observing $\rvy$, i.e., the conditional expectation,
\begin{eqnarray}
\label{eqn:g-MMSE}
g(y) = \mathbf{E}[\rvx | y].
\end{eqnarray}
The resulting maximized $\Delta_g$ and $I_{\mathrm{GMI}, g}$ are
\begin{eqnarray}
\label{eqn:delta-g-nonlinear}
\Delta_\mathrm{MMSE} = \frac{\mathrm{var} \mathbf{E}[\rvx | \rvy]}{P},
\end{eqnarray}
and
\begin{eqnarray}
\label{eqn:GMI-g-nonlinear}
I_{\mathrm{GMI}, \mathrm{MMSE}} = \frac{1}{2} \log \frac{P}{P - \mathrm{var} \mathbf{E}[\rvx | \rvy]},
\end{eqnarray}
respectively. The corresponding scaling parameter $a$ is exactly $\Delta_\mathrm{MMSE}$ in (\ref{eqn:delta-g-nonlinear}).
\end{prop}
{\it Proof:} In our information transmission model, we recognize the channel input $\rvx$ as $\rvu$ and the channel output $\rvy$ as $\rvv$ in Lemma \ref{lem:renyi-correlation-coefficient}. According to (\ref{eqn:correlation-coefficient-extremal}),
\begin{eqnarray}
\Theta_\rvy(\rvx) = \sup_g \left|\frac{\mathbf{E}[\rvx g(\rvy)]}{\sqrt{P \cdot \mathrm{var} g(\rvy)}}\right|,
\end{eqnarray}
noting that $\rvx \sim \mathcal{N}(0, P)$ under condition (1). Hence,
\begin{equation}
\label{eqn:21}
\Theta^2_\rvy(\rvx) = \sup_g \frac{\left(\mathbf{E}[\rvx g(\rvy)]\right)^2}{P \cdot \mathrm{var} g(\rvy)} \geq \sup_g \frac{\left(\mathbf{E}[\rvx g(\rvy)]\right)^2}{P \cdot \mathbf{E} [g(\rvy)^2]} = \sup_g \Delta_g.
\end{equation}
On the other hand, from Definition \ref{defn:renyi-correlation-coefficient},
\begin{eqnarray}
\Theta^2_\rvy(\rvx) = \frac{\mathrm{var} \mathbf{E}[\rvx | \rvy]}{\mathrm{var} \rvx} = \frac{\mathrm{var} \mathbf{E}[\rvx | \rvy]}{P}.
\end{eqnarray}
Therefore, (\ref{eqn:21}) becomes
\begin{eqnarray}
\sup_g \Delta_g \leq \frac{\mathrm{var} \mathbf{E}[\rvx | \rvy]}{P},
\end{eqnarray}
and equality can be attained, by letting $g(y) = \mathbf{E}[\rvx | y]$, because of Lemma \ref{lem:renyi-correlation-coefficient} and the fact that $\mathbf{E}\left[\mathbf{E}[\rvx | \rvy]\right] = \mathbf{E}[\rvx] = 0$. This establishes Proposition \ref{prop:nonlinear}.
{\bf Q.E.D.}

Here we provide a geometric interpretation of Proposition \ref{prop:nonlinear}. Inspecting the general expression of $\Delta_g$ in (\ref{eqn:gmi-generic}), we recognize it as the squared correlation coefficient between the channel input $\rvx$ and the processed channel output $g(\rvy)$, i.e., the squared cosine of the ``angle'' between $\rvx$ and $g(\rvy)$. So choosing the processing function $g$ means that we process the channel output $\rvy$ appropriately so as to ``align'' it with $\rvx$, and the best alignment is accomplished by the MMSE estimator, i.e., the conditional expectation operator.

Utilizing the orthogonality property of MMSE estimator, $\mathbf{E}\left[\left(\rvx - \mathbf{E}[\rvx | \rvy]\right) \mathbf{E}[\rvx | \rvy]\right] = 0$ (see, e.g., \cite{poor94:book}), we can verify that the denominator in the logarithm of $I_{\mathrm{GMI}, \mathrm{MMSE}}$ in (\ref{eqn:GMI-g-nonlinear}) is exactly the mean squared error of the MMSE estimator (\ref{eqn:g-MMSE}), which may be conveniently denoted by $\mathsf{mmse}$. So we have\footnote{A side note is that (\ref{eqn:GMI-g-MMSE}) is consistent with the so-called estimation counterpart to Fano's inequality \cite[Cor. of Thm. 8.6.6]{cover06:book}: $\mathbf{E}[(\rvx - \hat{\rvx}(\rvy))^2] \geq \frac{1}{2\pi e} e^{2 h(\rvx|\rvy)}$, where $\hat{\rvx}(\rvy)$ is an arbitrary estimate of $\rvx$ based upon $\rvy$. Under $\rvx \sim \mathcal{N}(0, P)$, we have $I(\rvx; \rvy) \geq I_{\mathrm{GMI}, \mathrm{MMSE}} = \frac{1}{2} \log \frac{P}{\mathsf{mmse}}$, i.e., $\mathsf{mmse} \geq P e^{-2 I(\rvx; \rvy)} = \frac{P}{e^{2 h(\rvx)}} e^{2 h(\rvx|\rvy)} = \frac{1}{2\pi e} e^{2 h(\rvx|\rvy)}$, thereby revisiting \cite[Cor. of Thm. 8.6.6]{cover06:book}.
}
\begin{eqnarray}
\label{eqn:GMI-g-MMSE}
I_{\mathrm{GMI}, \mathrm{MMSE}} = \frac{1}{2} \log \frac{P}{\mathsf{mmse}}.
\end{eqnarray}

\begin{figure}
\centering
\includegraphics[width=3.2in]{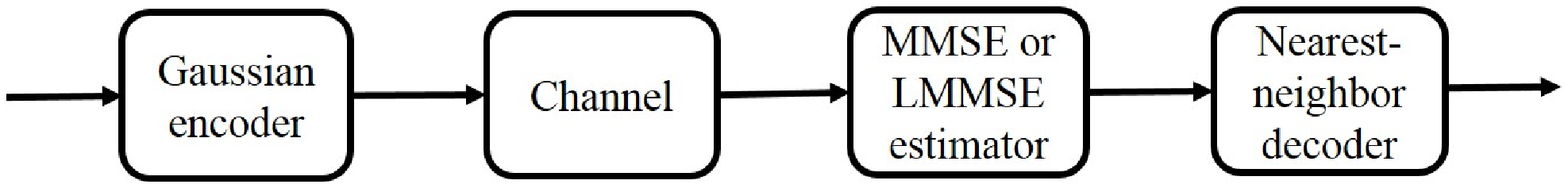}
\caption{Schematic diagram of transceiver structure.}
\label{fig:schematic}
\end{figure}

In Figure \ref{fig:schematic} we illustrate the transceiver structure suggested by Propositions \ref{prop:linear} and \ref{prop:nonlinear}. The key difference between these two propositions lies in the choice of the channel output processing function, and the effect is clearly seen by comparing (\ref{eqn:gmi-lmmse}) and (\ref{eqn:GMI-g-MMSE}). For certain channels, the performance of MMSE estimator may substantially outperform that of LMMSE estimator, and consequently the benefit in terms of GMI may be noticeable.

The data processing inequality asserts that for any channel, processing the channel output cannot increase the input-output mutual information \cite{cover06:book}, but Propositions \ref{prop:linear} and \ref{prop:nonlinear} do not violate it. This is because in our information transmission model, the decoder structure is restricted to be a nearest neighbor rule, which may be mismatched to the channel.

In order to illustrate the analytical results in this section, we present in Appendix case studies about a single-input-multiple-output (SIMO) channel without output quantization and with one-bit output quantization (with and without dithering). These channel models will also be used as examples in our study of learning based information transmission, in the next section.

\section{Learning Based Information Transmission}
\label{sec:learning}

With the key insights gained in Section \ref{sec:memoryless}, in this section we turn to the setting where the encoder and the decoder do not utilize the statistical channel model, and study how to incorporate machine learning ingredients into our problem.

\subsection{Analogy with Regression Problems}
\label{subsec:relationship-regression}

From the study in Section \ref{sec:memoryless}, we see that with the codebook ensemble and the decoder structure fixed as in conditions (1) and (2), the key task is to choose an appropriate processing function so as to ``align'' the processed channel output $g(\rvy)$ with the channel input $\rvx$. When the statistical channel model is known, the optimal choice of $g$ is the conditional expectation operator. However, without utilizing the channel model knowledge, we need to learn a ``good'' choice of $g$ based on training data. This is where the theory of machine learning kicks in.

Our problem is closely related to the classical topic of regression in machine learning. In regression, we need to predict the value of a random variable $\rvx$ upon observing another random variable $\rvy$.\footnote{In most machine learning literatures (see, e.g., \cite{hastie09:book}), $\rvy$ is used for representing the quantity to predict and $\rvx$ for the observed, exactly in contrary to our convention here. The reason for adopting our convention is that for information transmission problems $\rvx$ is used for representing channel input and $\rvy$ for channel output.} Under quadratic loss, if the statistics of $(\rvx, \rvy)$ is known, then the optimal regression function is the conditional expectation operator. In the absence of statistical model knowledge, extensive studies have been devoted to design and analysis of regression functions that behave similarly to the conditional expectation operator; see, e.g., \cite{hastie09:book}.

We note that, although our problem and the classical regression problem both boil down to designing processing functions that are in some sense ``close'' to the conditional expectation operator, there still exist some fundamental distinctions between the two problems. In short, the distinctions are due to the different purposes of the two problems. For a regression problem, we assess the quality of a predictor through its generalization error, which is the expected loss when applying the predictor to a new pair of $(\rvx, \rvy)$, besides the training data set \cite[Chap. 7]{hastie09:book}. For our information transmission problem, from a training data set, we not only need to form a processing function, but also need to determine a rate for transmitting messages. So the code rate is not a priori known, but need be training data dependent. We assess the quality of our design through certain characteristics of the rate. The details of our problem formulation are in the next subsection.

\subsection{Problem Formulation}
\label{subsec:formulation}

Before the information transmission phase, we have a learning phase. Suppose that we are given $L$ i.i.d. pairs of $(\rvx, \rvy)$ as the training data, according to the channel input-output joint probability distribution $p(x, y) = p(x) p(y|x)$, which we denote by
\begin{eqnarray}
\mathcal{T} = \left\{(\rvx_1, \rvy_1), \ldots, (\rvx_L, \rvy_L)\right\}.
\end{eqnarray}
We have two tasks in the learning phase, with the training data set $\mathcal{T}$. First, we need to form a processing function $g_\mathcal{T}$ and a scaling parameter $a_\mathcal{T}$, which will be used by the decoder to implement its decoding rule. Second, we need to provide a code rate $R_\mathcal{T}$ so that the encoder and the decoder can choose their codebook to use during the information transmission phase. According to our discussion in Section \ref{subsec:relationship-regression}, we desire to make $g_\mathcal{T}$ close to the conditional expectation operator.

From a design perspective, it makes sense to distinguish two possible scenarios:

(A) $p(y|x)$ is unknown, and

(B) $p(y|x)$ is too complicated to prefer an exact realization of $g(y) = \mathbf{E}[\rvx|y]$, but is still known to the decoder so that it can simulate the channel accordingly.

To justify scenario (B), we note that for a given $p(x)p(y|x)$, it is usually easy to generate a random channel output $\rvy$ for any given channel input $x$ according to $p(y|x)$, but very difficult to inversely compute $\mathbf{E}[\rvx|y]$ for a given channel output $y$ because that generally involves marginalization which can be computationally intensive for high-dimensional $\mathcal{Y}$.

We generate the training data set $\mathcal{T}$ as follows. Under scenario (A), the encoder transmits i.i.d. training inputs $\rvx_1, \ldots, \rvx_L$, known to the decoder in advance, through the channel to the decoder, and the decoder thus obtains $\mathcal{T}$. Note that since we have control over the encoder, the input distribution $p(x)$ is known. In contrast, under scenario (B), no actual transmission of training inputs is required, and the decoder simulates the channel by himself, according to $p(x) p(y|x)$, in an offline fashion, to obtain $\mathcal{T}$. We emphasize that, here in the learning phase, the input distribution $p(x)$ need not be the same as that in the information transmission phase (i.e., Gaussian). Changing $p(x)$ certainly will change the distribution of $\mathcal{T}$, correspondingly the regression performance, and eventually the information transmission performance. The Gaussian distribution does not necessarily bear any optimality for a general statistical channel model. Nevertheless, in developing our proposed algorithm in Section \ref{subsec:algorithm} and conducting  numerical experiments in Section \ref{subsec:learning-case-study}, we require the training inputs be Gaussian to generate $\mathcal{T}$, for technical reasons.

\begin{figure}
\centering
\includegraphics[width=3.2in]{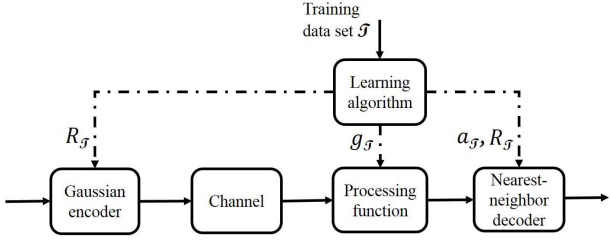}
\caption{Schematic diagram of learning based transceiver structure.}
\label{fig:learning-schematic}
\end{figure}

It is always the decoder who accomplishes the learning task aforementioned, and informs the encoder the value of $R_\mathcal{T}$, possibly via a low-rate control link. In Figure \ref{fig:learning-schematic} we illustrate the transceiver structure when the learning phase is taken into consideration.

Under scenario (A), we learn both $g_\mathcal{T}$ and $a_\mathcal{T}$; while under scenario (B), we learn $g_\mathcal{T}$, and can subsequently calculate the corresponding optimal scaling parameter $a = \mathbf{E}[\rvx g_\mathcal{T}(\rvy)]/P$, based upon Proposition \ref{prop:generic-g}. More details about how learning is accomplished will be given in the later part of this subsection and in Section \ref{subsec:algorithm}. The achieved GMIs under the two scenarios are different, as given by the following result.

\begin{prop}
\label{prop:gmi-gT}
Consider the information transmission model under conditions (1) and (2), given a training data set $\mathcal{T}$ and a certain learning algorithm.

Under scenario (A), denote the learnt $(g, a)$ pair by $(g_\mathcal{T}, a_\mathcal{T})$. The corresponding GMI is given by

\begin{eqnarray}
\begin{split}
\label{eqn:gmi-gT-optJ}
I_{\mathrm{GMI}, \mathcal{T}, A} = \max_{\gamma \geq 0} \bigg \{& \frac{1}{2} \log(1 + \gamma) - \frac{\gamma}{2} - \frac{\gamma^2}{1 + \gamma} \frac{\mathbf{E}\left[g_\mathcal{T}(\rvy)^2\right]}{2 a_\mathcal{T}^2 P} \\
& + \gamma \frac{\mathbf{E}[\rvx g_\mathcal{T}(\rvy)]}{a_\mathcal{T} P}\bigg\}.
\end{split}
\end{eqnarray}

Under scenario (B), denote the learnt $g$ by $g_\mathcal{T}$. The corresponding GMI is given by
\begin{eqnarray}
\label{eqn:gmi-scenario-B}
I_{\mathrm{GMI}, \mathcal{T}, B} = \frac{1}{2} \log \frac{1}{1 - \Delta_{\mathcal{T}, B}},
\end{eqnarray}
where $\Delta_{\mathcal{T}, B} = \frac{\left(\mathbf{E}[\rvx g_\mathcal{T}(\rvy)]\right)^2}{P \mathbf{E}[g_\mathcal{T}(\rvy)^2]}.$ In both (\ref{eqn:gmi-gT-optJ}) and (\ref{eqn:gmi-scenario-B}), the expectations are evaluated under $\rvx \sim \mathcal{N}(0, P)$.
\end{prop}
{\it Proof:} Consider scenario (A). We still follow the proof of Proposition \ref{prop:generic-g}, viewing $(g_\mathcal{T}, a_\mathcal{T})$ as a specific choice in the decoding rule. With a fixed $a_\mathcal{T}$, the maximization problem (\ref{eqn:gmi-maximizeJ}) is with respect to $\theta$ only; that is,
\begin{eqnarray}
\begin{split}
I_{\mathrm{GMI}, \mathcal{T}, A} = \max_{\theta < 0} \bigg\{&\theta \mathbf{E}\left[\left(g_\mathcal{T}(\rvy) - a_\mathcal{T} \rvx \right)^2\right] - \frac{\theta \mathbf{E}[g_\mathcal{T}(\rvy)^2]}{1 - 2 \theta a_\mathcal{T}^2 P} \\
&+ \frac{1}{2} \log(1 - 2\theta a_\mathcal{T}^2 P)\bigg \}.
\end{split}
\end{eqnarray}
Rearranging terms, and making a change of variable $\gamma = - 2\theta a_\mathcal{T}^2 P > 0$, we obtain (\ref{eqn:gmi-gT-optJ}).

Consider scenario (B), where the decoder knows the statistical channel model $p(y|x)$. Therefore, according to Proposition \ref{prop:generic-g}, upon learning $g_\mathcal{T}$, he can set the optimal choice of the scaling parameter $a$ as $\mathbf{E}[\rvx g_\mathcal{T}(\rvy)]/P$, resulting in $\Delta_{\mathcal{T}, B}$ and $I_{\mathrm{GMI}, \mathcal{T}, B}$ in (\ref{eqn:gmi-scenario-B}). {\bf Q.E.D.}

It is clear that $I_{\mathrm{GMI}, \mathcal{T}, A}$ is no greater than $I_{\mathrm{GMI}, \mathcal{T}, B}$, and their gap is due to the lack of the statistical channel model knowledge $p(y|x)$. It is natural to expect that when learning is effective, the gap will be small. The following corollary illustrates one such case.

\begin{cor}
\label{cor:AB-gap}
Suppose that a learning algorithm learns $g_\mathcal{T}$ under both scenarios (A) and (B), and under scenario (A) also learns $a_\mathcal{T}$. Denote the gap between
\begin{eqnarray}
\Delta_{\mathcal{T}, A} = \frac{a_\mathcal{T} \mathbf{E}\left[\rvx g_\mathcal{T}(\rvy)\right]}{\mathbf{E}\left[g_\mathcal{T}(\rvy)^2\right]}\quad\mbox{and}\;
\Delta_{\mathcal{T}, B} = \frac{\left(\mathbf{E}[\rvx g_\mathcal{T}(\rvy)]\right)^2}{P \mathbf{E}[g_\mathcal{T}(\rvy)^2]}
\end{eqnarray}
by $\delta = \Delta_{\mathcal{T}, A} - \Delta_{\mathcal{T}, B}$. When $a_\mathcal{T}$ satisfies $0 < a_\mathcal{T} \mathbf{E}\left[\rvx g_\mathcal{T}(\rvy)\right] < \mathbf{E}\left[g_\mathcal{T}(\rvy)^2\right]$, we have $I_{\mathrm{GMI}, \mathcal{T}, B} = I_{\mathrm{GMI}, \mathcal{T}, A} + O(\delta^2)$, i.e., the gap between the two GMIs is quadratic with $\delta$.
\end{cor}
{\it Proof:}
Under the condition for $a_\mathcal{T}$, we have $0 < \Delta_\mathcal{T} < 1$, and we can choose a specific $\gamma = \Delta_\mathcal{T}/(1 - \Delta_\mathcal{T}) > 0$ in (\ref{eqn:gmi-lb-T}) to get a lower bound of $I_{\mathrm{GMI}, \mathcal{T}, A}$ as
\begin{equation}\label{eqn:gmi-lb-T}
I_{\mathrm{GMI}, \mathcal{T}, A} \geq \underline{I}_{\mathrm{GMI}, \mathcal{T}, A} = \frac{1}{2} \log \frac{1}{1 - \Delta_{\mathcal{T}, A}} - \frac{\Delta_{\mathcal{T}, A} - \Delta_{\mathcal{T}, B}}{2\left(1 - \Delta_{\mathcal{T}, A}\right)}.
\end{equation}
Via a Taylor expansion with respect to $\delta = \Delta_{\mathcal{T}, A} - \Delta_{\mathcal{T}, B}$, we have that (\ref{eqn:gmi-lb-T}) behaves like $\underline{I}_{\mathrm{GMI}, \mathcal{T}, A} = I_{\mathrm{GMI}, \mathcal{T}, B} - O(\delta^2)$, where $O(\delta^2)/\delta^2$ is bounded as $\delta \rightarrow 0$. Therefore, the gap between $I_{\mathrm{GMI}, \mathcal{T}, A}$ and $I_{\mathrm{GMI}, \mathcal{T}, B}$ is $O(\delta^2)$.
{\bf Q.E.D.}

Under scenario (A), a learning algorithm is a mapping $\mathcal{T} \mapsto (g_\mathcal{T}, a_\mathcal{T}, R_\mathcal{T})$. The resulting output processing function (called regression function or predictor in classical regression problems) $g_\mathcal{T}: \mathcal{Y} \mapsto \mathcal{X} = \mathbb{R}$ usually belongs to certain prescribed function class, which may be linear (e.g., least squares, ridge regression) or nonlinear (e.g., kernel smoothing, neural networks). According to Proposition \ref{prop:gmi-gT}, we should set $R_\mathcal{T} = I_{\mathrm{GMI}, \mathcal{T}, A}$. This is, however, impossible since neither the encoder nor the decoder can calculate (\ref{eqn:gmi-gT-optJ}), without knowing $p(y|x)$. The situation is that the rate $I_{\mathrm{GMI}, \mathcal{T}, A}$ is achievable but its value is ``hidden'' by the nature. We hence need to estimate it, again based on $\mathcal{T}$ and its induced $(g_\mathcal{T}, a_\mathcal{T})$. We desire a highly asymmetric estimation; that is, $R_\mathcal{T} \leq I_{\mathrm{GMI}, \mathcal{T}}$ should hold with high probability, since otherwise there would be no guarantee on the achievability of $R_\mathcal{T}$, resulting in decoding failures. Meanwhile, we also desire that $R_\mathcal{T}$ is close to $I_{\mathrm{GMI}, \mathrm{MMSE}}$, which corresponds to the ideal situation where the statistical channel model $p(y|x)$ is known to the encoder and the decoder.

The learnt $(g_\mathcal{T}, a_\mathcal{T}, R_\mathcal{T})$ are random due to the randomness of $\mathcal{T}$. In order to assess the performance of learning, based upon our discussion, we introduce the following two metrics to quantify the learning loss:
\begin{itemize}
\item Over-estimation probability:
\begin{eqnarray}
\label{eqn:oep}
\mathcal{P}_\mathrm{oe} = \mathbf{Pr}\left[R_\mathcal{T} > I_{\mathrm{GMI}, \mathcal{T}, A}\right].
\end{eqnarray}
This may be understood as the ``outage probability'' corresponding to a learning algorithm.
\item Receding level:
\begin{eqnarray}
\label{eqn:rl}
\mathcal{L}_\mathrm{r} = 1 - \frac{\mathbf{E}\left[R_\mathcal{T}|I_{\mathrm{GMI}, \mathcal{T}, A}- R_\mathcal{T} \geq 0\right]}{I_{\mathrm{GMI, MMSE}}}.
\end{eqnarray}
This is the averaged relative gap between the learnt code rate $R_\mathcal{T}$ and the GMI under known channel and optimal output processing, conditioned upon the event that over-estimation does not occur.
\end{itemize}
It is certainly desirable to have both $\mathcal{P}_\mathrm{oe}$ and $\mathcal{L}_\mathrm{r}$ small.

Under scenario (B), the situation is much simpler. A learning algorithm is a mapping $\mathcal{T} \mapsto g_\mathcal{T}$, and based upon $g_\mathcal{T}$ we can choose $R_\mathcal{T} = I_{\mathrm{GMI}, \mathcal{T}, B}$ since it can be evaluated as shown in Proposition \ref{prop:gmi-gT}. So we do not need to consider over-estimation, and the receding level is simply
\begin{eqnarray}\label{eqn:rl-B}
\mathcal{L}_\mathrm{r} = 1 - \frac{\mathbf{E}\left[I_{\mathrm{GMI}, \mathcal{T}, B}\right]}{I_{\mathrm{GMI}, \mathrm{MMSE}}}.
\end{eqnarray}

\begin{figure}
    \centering
        \begin{minipage}{8cm}
        \includegraphics[width=3.in]{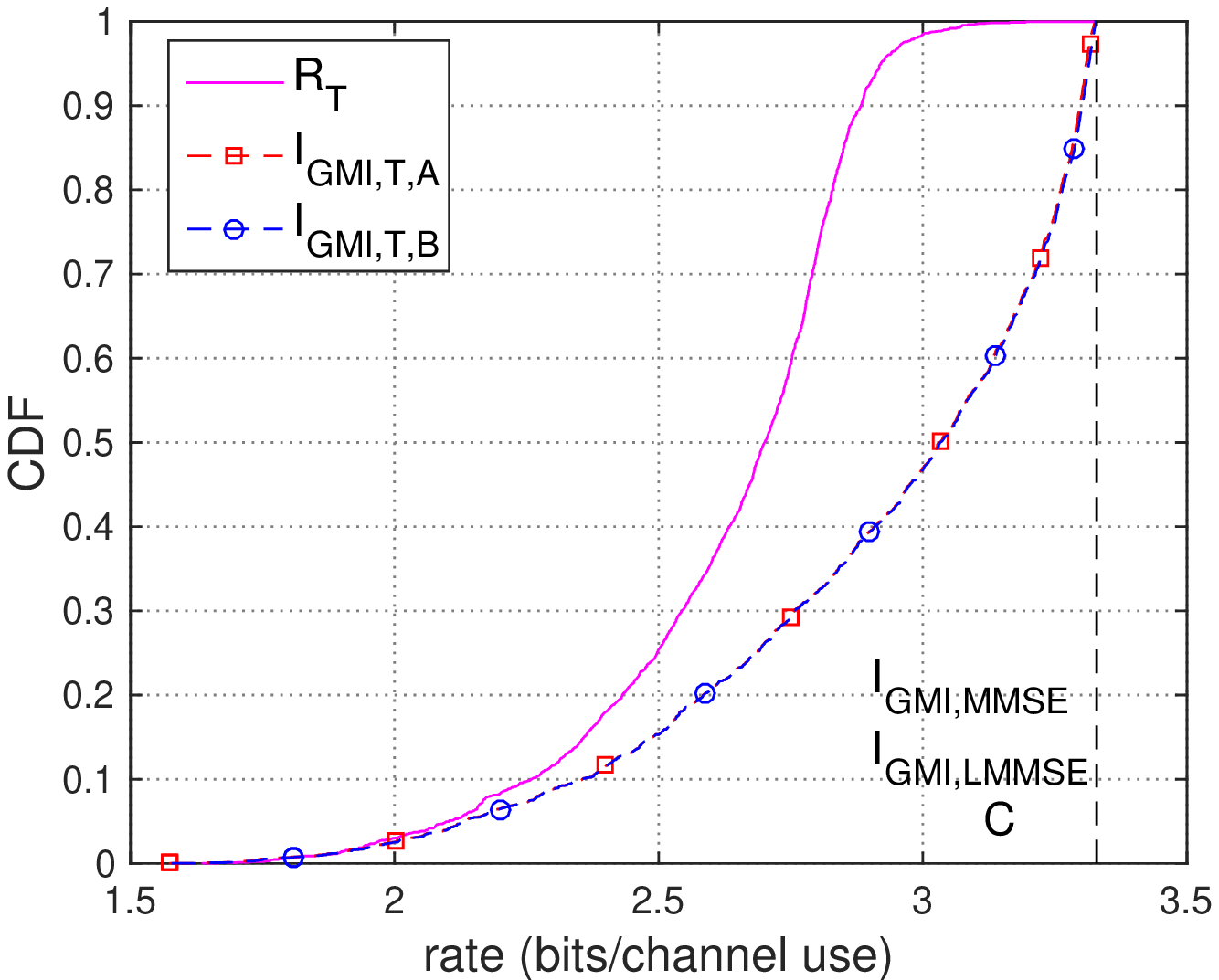}
        \caption{An AWGN example: CDFs of rates.}
        \label{fig:AWGN-rates}
        \end{minipage}
        \begin{minipage}{8cm}
        \includegraphics[width=3.in]{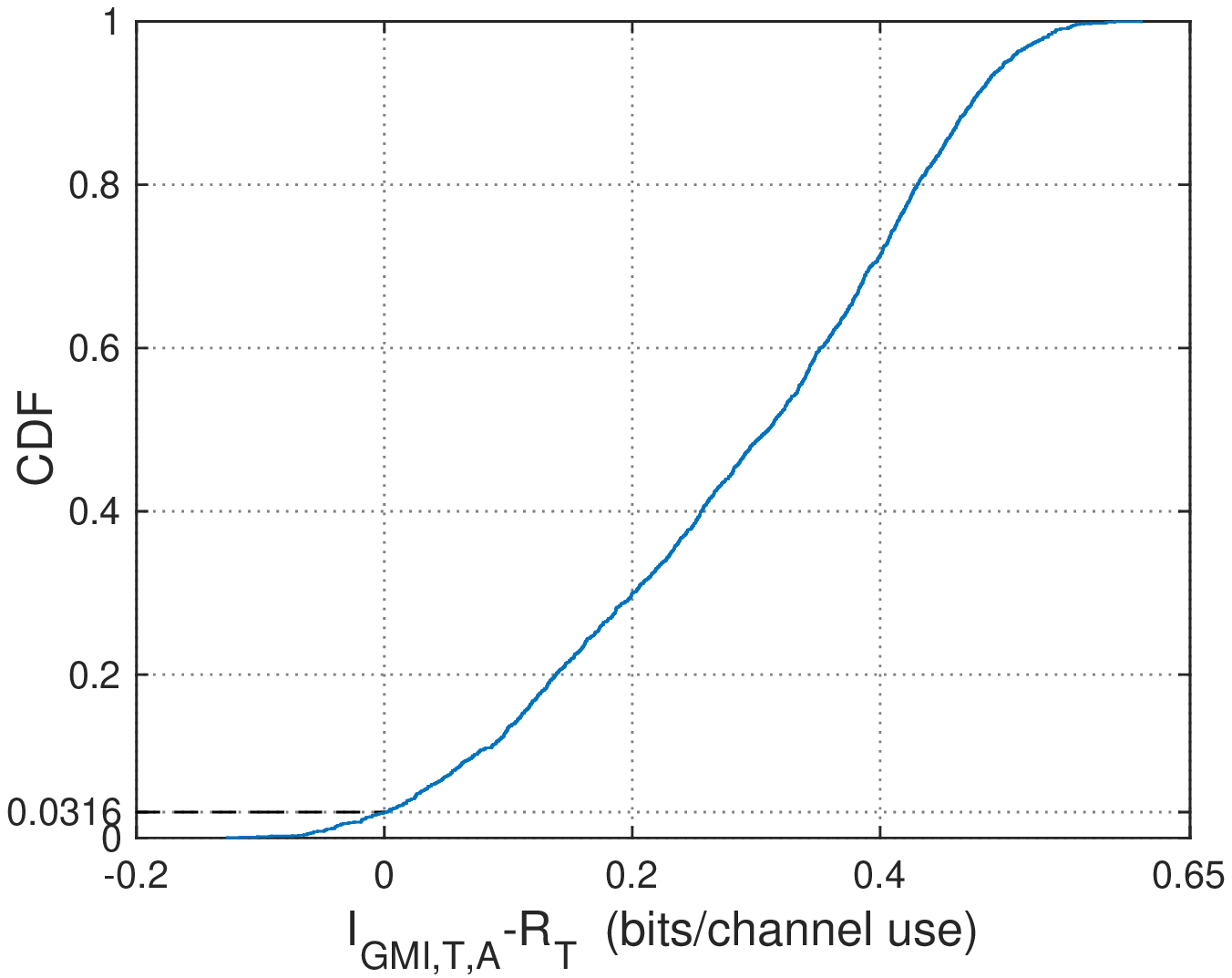}
        \caption{An AWGN example: illustration of over-estimation.}
        \label{fig:AWGN-metrics}
        \end{minipage}
\end{figure}

We illustrate the rates and loss metrics using a simple example of additive white Gaussian noise (AWGN) channel, $\rvy = \rvx + \rvz$, $\rvz \sim \mathcal{N}(0, 1)$, and $\mathbf{E}[\rvx^2] \leq P = 100$. We use the {\sf LFIT} algorithm proposed in Section \ref{subsec:algorithm} to accomplish the learning task. Figure \ref{fig:AWGN-rates} displays the cumulative distribution functions (CDFs) of the resulting $I_{\mathrm{GMI}, \mathcal{T}, A}$, $I_{\mathrm{GMI}, \mathcal{T}, B}$, and $R_\mathcal{T}$. As suggested by Corollary \ref{cor:AB-gap}, the gap between $I_{\mathrm{GMI}, \mathcal{T}, A}$ and $I_{\mathrm{GMI}, \mathcal{T}, B}$ is nearly negligible. Note that for AWGN channels, $I_{\mathrm{GMI}, \mathrm{MMSE}} = I_{\mathrm{GMI}, \mathrm{LMMSE}}$ and is further equal to channel capacity $(1/2) \log (1 + P)$ (the dashed vertical line). Figure \ref{fig:AWGN-metrics} displays the CDF of $\left(I_{\mathrm{GMI}, \mathcal{T}, A} - R_\mathcal{T}\right)$, so the negative part corresponds to over-estimation events and the y-intercept is $\mathcal{P}_\mathrm{oe}$ (3.16\% in this example).

In Table I, we summarize a comparison among the problem formulations of the two scenarios we consider and classical regression problems (see, e.g., \cite{hastie09:book}).

\begin{table*}[h!]
\label{table:comparison}
\begin{center}
\caption{Comparison of problem formulations}
\begin{tabular}{l|l|l|l}
& Classical regression & Scenario (A) & Scenario (B)\\
\hline
Learning algorithm & $\mathcal{T} \mapsto g_\mathcal{T}$ & $\mathcal{T} \mapsto (g_\mathcal{T}, a_\mathcal{T}, R_\mathcal{T})$ & $\mathcal{T} \mapsto g_\mathcal{T}$ \\
\hline
Processing function $g_\mathcal{T}$ & Regression function, linear or nonlinear & \multicolumn{2}{|l}{Output processing function, linear or nonlinear} \\
\hline
Ground truth of $g_\mathcal{T}$ & \multicolumn{3}{|c}{MMSE estimator $g(y) = \mathbf{E}[\rvx |y]$}\\
\hline
Loss function & $\mathbf{E}\left[(\rvx - g_\mathcal{T}(\rvy))^2\right]$ & $\mathcal{P}_\mathrm{oe}$ (\ref{eqn:oep}) and $\mathcal{L}_\mathrm{r}$ (\ref{eqn:rl}) & $\mathcal{L}_\mathrm{r}$ (\ref{eqn:rl-B}) \\
\hline
\end{tabular}
\end{center}
\end{table*}

\subsection{Proposed Algorithm}
\label{subsec:algorithm}

There already exist numerous learning algorithms for classical regression problems to obtain $g_\mathcal{T}$, but under scenario (A) we further need to obtain $a_\mathcal{T}$ and $R_\mathcal{T}$. Meanwhile, the learning loss measured by over-estimation probability and receding level are also unconventional. We study these in this subsection.

We begin with a sketch of our approach. We use a collection of parameters $\lambda$ to specify the structure used by $g_\mathcal{T}$. The exact choice of $\lambda$ will be discussed in Section \ref{subsec:learning-case-study}, which can be, for example, the complexity parameter in ridge regression, the width of kernel in kernel smoothing methods, the hyperparameters in deep neural networks, and so on. Fixing $\lambda$, based upon $\mathcal{T}$, we learn $g_\mathcal{T}$, and then on top of it, estimate $a_\mathcal{T}$ and $R_\mathcal{T}$.

We can provide a theoretical guarantee on the over-estimation probability, as given by the following result, whose proof technique will also be useful for devising our proposed algorithm.

\begin{prop}
\label{prop:Poe-guarantee}
Suppose that we have generated a length-$L$ training data set $\mathcal{T}$ subject to $\rvx \sim \mathcal{N}(0, P)$. We split it into two parts, of lengths $(1 - \nu) L$ and $\nu L$, respectively, for some $\nu \in (0, 1)$, and we learn $g_\mathcal{T}$ and $a_\mathcal{T}$ solely based upon the length-$(1 - \nu) L$ part. Then as $L \rightarrow \infty$, the following estimate of $R_\mathcal{T}$ achieves an over-estimation probability no greater than $\mathcal{P}_\mathrm{oe}$:
\begin{equation}
\label{eqn:rate-Poe-guarantee}
R_\mathcal{T} = \max_{\gamma \geq 0} \left\{\frac{1}{2} \log(1 + \gamma) - \frac{\gamma}{2} - \frac{\gamma^2}{1 + \gamma} F_\rvy + \gamma F_{\rvx \rvy}\right\},
\end{equation}
\begin{equation}
\label{eqn:Fy}
F_\rvy = \frac{1}{2 a_\mathcal{T}^2 P} \left[\frac{1}{\nu L}\sum_{i = 1}^{\nu L} g_\mathcal{T}(\rvy_i)^2 + \frac{\sqrt{2\overline{\mathrm{var}}\left(g_\mathcal{T}(\rvy)^2\right)}}{\sqrt{\nu L}}\mathrm{erfc}^{-1}(\mathcal{P}_\mathrm{oe})\right],\\
\end{equation}
\begin{equation}
\begin{split}
\label{eqn:Fxy}
&F_{\rvx \rvy} = \\
&\frac{1}{a_\mathcal{T} P} \left[\frac{1}{\nu L}\sum_{i = 1}^{\nu L} \rvx_i g_\mathcal{T}(\rvy_i) - \mathrm{sgn}(a_\mathcal{T}) \frac{\sqrt{2\overline{\mathrm{var}}\left(\rvx g_\mathcal{T}(\rvy)\right)}}{\sqrt{\nu L}}\mathrm{erfc}^{-1}(\mathcal{P}_\mathrm{oe})\right].
\end{split}
\end{equation}
Here $\overline{\mathrm{var}} (U)$ denotes the empirical variance for i.i.d. random variables $U_1, U_2, \ldots, U_{\nu L}$, i.e.,
\begin{eqnarray}
\overline{\mathrm{var}} (U) = \frac{1}{\nu L - 1} \sum_{i = 1}^{\nu L} \left[U_i - \frac{1}{\nu L}\sum_{j = 1}^{\nu L} U_j\right]^2.
\end{eqnarray}
\end{prop}
{\it Proof:} We start with the expression of $I_{\mathrm{GMI}, \mathcal{T}, A}$ (\ref{eqn:gmi-gT-optJ}) in Proposition \ref{prop:gmi-gT}. Since $\gamma \geq 0$, it is clear that the maximum value of the right hand side of (\ref{eqn:gmi-gT-optJ}) will exceed $I_{\mathrm{GMI}, \mathcal{T}, A}$ only if (i) the estimate of $\mathbf{E}\left[g_\mathcal{T}(\rvy)^2\right]$ is smaller than its true value, or (ii) the estimate of $\mathbf{E}\left[\rvx g_\mathcal{T}(\rvy)\right]$ is larger (smaller) than its true value when $a_\mathcal{T}$ is positive (negative). Therefore, in order to ensure an over-estimation probability target $\mathcal{P}_\mathrm{oe}$, it suffices to require each of (i) and (ii) occurs with probability no larger than $\mathcal{P}_\mathrm{oe}/2$, due to the union bound. Applying the central limit theorem (CLT) \cite[Thm. 27.1]{billingsley:book} thus leads to (\ref{eqn:Fy}) and (\ref{eqn:Fxy}), which ensure that $R_\mathcal{T}$ in (\ref{eqn:rate-Poe-guarantee}) does not exceed $I_{\mathrm{GMI}, \mathcal{T}, A}$ in (\ref{eqn:gmi-gT-optJ}) with probability no smaller than $1 - \mathcal{P}_\mathrm{oe}$, as $L \rightarrow \infty$. {\bf Q.E.D.}

We remark that, by imposing appropriate regularity conditions, we may replace the bias terms in (\ref{eqn:Fy}) and (\ref{eqn:Fxy}) using results from concentration inequalities (e.g., Bernstein's inequalities \cite[Thm. 2.10]{boucheron:book}), which control the over-estimation probability even for finite $L$. In practice, we find that while both CLT and concentration inequalities control the over-estimation probability well below its target, they lead to rather large receding level, unless the training data set size is extremely large. This appears to be due to that the bias terms in (\ref{eqn:Fy}) and (\ref{eqn:Fxy}) tend to be overly conservative when plugged into the maximization (\ref{eqn:rate-Poe-guarantee}). For this reason, in the following, we propose an algorithm to produce $(g_\mathcal{T}, a_\mathcal{T}, R_\mathcal{T})$, which has better performance in numerical experiments, applying the idea of cross-validation (CV) \cite[Chap. 7, Sec. 10]{hastie09:book}, and drawing insights from Proposition \ref{prop:Poe-guarantee}.

The motivation of applying CV is as follows. For any reasonable size $L$ of $\mathcal{T}$, we need to utilize the training data economically. But if we simply use the same training data for both learning $g_\mathcal{T}$ and estimating $a_\mathcal{T}$ and $R_\mathcal{T}$ (which involve $g_\mathcal{T}$), a delicate issue is that the resulting estimates will usually be severely biased, a phenomenon already well known in classical regression problems when assessing generalization error using training data \cite[Chap. 7, Sec. 2]{hastie09:book}. CV is an effective approach for alleviating such a problem, and the procedure is as follows.

We split $\mathcal{T}$ into $Q$ non-overlapping segments of the same size, indexed from $1$ to $Q$. Taking away the $q$-th segment, we learn $g_\mathcal{T}$ using the remaining $Q - 1$ segments, and then use the $q$-th segment to estimate expectations needed for calculating $a_\mathcal{T}$ and $R_\mathcal{T}$. As $q$ runs from $1$ to $Q$, we have $Q$ estimates for each interested expectation, and average them as the final estimate.

Denote the training data in the $q$-th segment as $\mathcal{T}_q = \left\{(\rvx_1^{(q)}, \rvy_1^{(q)}), \ldots, (\rvx_{L/Q}^{(q)}, \rvy_{L/Q}^{(q)})\right\}$, and the learnt $g_\mathcal{T}$ without using $\mathcal{T}_q$ as $g_{\mathcal{T}, -q}$. Note that $\mathcal{T}_q$, $q = 1, \ldots, Q$, are disjoint and $\mathcal{T} = \bigcup_{q = 1}^Q \mathcal{T}_q$.

With $g_{\mathcal{T}, -q}$ and $\mathcal{T}_q$, we estimate the following expectations via empirical means:
\begin{eqnarray}\label{eqn:empirical-mean-q}
\begin{split}
\hat{\mathbf{E}}[\rvx g_{\mathcal{T}, -q}(\rvy)] &= \frac{Q}{L} \sum_{l = 1}^{L/Q} \rvx_l^{(q)} g_{\mathcal{T}, -q} \left(\rvy_l^{(q)}\right), \\
\hat{\mathbf{E}}[g_{\mathcal{T}, -q}(\rvy)^2] &= \frac{Q}{L} \sum_{l = 1}^{L/Q} g_{\mathcal{T}, -q} \left(\rvy_l^{(q)}\right)^2.
\end{split}
\end{eqnarray}
Then we average them from $q = 1$ to $Q$, to obtain their CV estimates:
\begin{eqnarray}\label{eqn:empirical-mean-CV}
\begin{split}
\hat{\mathbf{E}}[\rvx g_\mathcal{T}(\rvy)] &= \frac{1}{Q} \sum_{q = 1}^Q \hat{\mathbf{E}}[\rvx g_{\mathcal{T}, -q}(\rvy)],\\
\hat{\mathbf{E}}[g_\mathcal{T}(\rvy)^2] &= \frac{1}{Q} \sum_{q = 1}^Q \hat{\mathbf{E}}[g_{\mathcal{T}, -q}(\rvy)^2].
\end{split}
\end{eqnarray}
We also use an empirical mean estimate of the second moment of $\rvx$, $P$, as $\hat{P} = \frac{1}{L} \sum_{l = 1}^L \rvx_l^2$. Based upon these, we choose $a_\mathcal{T}$ as $a_\mathcal{T} = \hat{\mathbf{E}}[\rvx g_\mathcal{T}(\rvy)]/\hat{P}$.

Now according to the proof of Proposition \ref{prop:Poe-guarantee}, we can affect the over-estimation probability by biasing the estimates $\hat{\mathbf{E}}[\rvx g_\mathcal{T}(\rvy)]$ and $\hat{\mathbf{E}}[g_\mathcal{T}(\rvy)^2]$. To implement this, we introduce two tunable scaling parameters which are typically close to one, in order to slightly bias these expectation estimates; that is, for prescribed $\xi_1 > 1$, $\xi_2 < 1$ if $a_\mathcal{T} > 0$ and $\xi_2 > 1$ otherwise, we choose the code rate $R_\mathcal{T}$ according to
\begin{equation}
\begin{split}
\label{eqn:R_T-scaling}
R_\mathcal{T} = \max_{\gamma \geq 0} \bigg\{&\frac{1}{2} \log(1 + \gamma) - \frac{\gamma}{2} - \frac{\gamma^2}{1 + \gamma} \frac{\xi_1 \hat{\mathbf{E}}[g_\mathcal{T}(\rvy)^2]}{2a_\mathcal{T}^2 \hat{P}} \\
&+ \gamma \frac{\xi_2 \hat{\mathbf{E}}[\rvx g_\mathcal{T}(\rvy)]}{a_\mathcal{T} \hat{P}}\bigg\}.
\end{split}
\end{equation}

We summarize the above ideas in the following algorithm.

\hrulefill

Algorithm $\mathsf{LFIT}$ (Learn For Information Transmission)

Input: $\mathcal{T}$, $\lambda$, $Q$, $\xi_1$, $\xi_2$.

Output: $g_\mathcal{T}$, $a_\mathcal{T}$, $R_\mathcal{T}$.

\begin{enumerate}
\item Partition $\mathcal{T}$ into $\mathcal{T}_1, \ldots, \mathcal{T}_Q$.
\item For $q$ from $1$ to $Q$:
\begin{enumerate}
\item[2a)] Obtain $g_{\mathcal{T}, -q}$ according to the specified processing function structure with parameter $\lambda$.
\item[2b)] Compute $\hat{\mathbf{E}}[\rvx g_{\mathcal{T}, -q}(\rvy)]$ and $\hat{\mathbf{E}}[g_{\mathcal{T}, -q}(\rvy)^2]$ according to (\ref{eqn:empirical-mean-q}).
\end{enumerate}
\item Compute $\hat{\mathbf{E}}[\rvx g_\mathcal{T}(\rvy)]$ and $\hat{\mathbf{E}}[g_\mathcal{T}(\rvy)^2]$ according to (\ref{eqn:empirical-mean-CV}), and $\hat{P} = \frac{1}{L} \sum_{l = 1}^L \rvx_l^2$.
\item Compute and output $g_\mathcal{T} = \frac{1}{Q} \sum_{q = 1}^Q g_{\mathcal{T}, -q}$.
\item Compute and output $a_\mathcal{T} = \hat{\mathbf{E}}[\rvx g_\mathcal{T}(\rvy)]/\hat{P}$.
\item Compute and output $R_\mathcal{T}$ according to (\ref{eqn:R_T-scaling}).
\end{enumerate}

\hrulefill

\subsection{Case Studies}
\label{subsec:learning-case-study}

A theoretical analysis of the $\mathsf{LFIT}$ algorithm in terms of $\mathcal{P}_\mathrm{oe}$ and $\mathcal{L}_\mathrm{r}$ appears to be elusive and is left for future research. In this subsection, we present numerical experiments with two representative types of processing functions, namely, ridge regression which is linear, and kernel smoother which is nonlinear. More complicated processing functions such as neural networks are left for future research. Note that the numerical experiments are all under scenario (A).

We consider the three channel models in the appendix, i.e., SIMO channel without quantization, with one-bit quantization, and with dithered one-bit quantization. For simplicity we consider the real case only. In our numerical experiments, we always let $\rvx \sim \mathcal{N}(0, P)$.

For ridge regression, with a training data set $\mathcal{T} = \{(\rvx_1, \rvy_1), \ldots, (\rvx_L, \rvy_L)\}$, the processing function is (see, e.g., \cite[Chap. 3, Sec. 4]{hastie09:book})
\begin{eqnarray}
\begin{split}
g(y) = y^T (\underline{\rvy}^T \underline{\rvy} + \lambda \mathbf{I}_p)^{-1} \underline{\rvy}^T \underline{\rvx}, \\
\underline{\rvx} = [\rvx_1, \ldots, \rvx_L]^T, \quad \underline{\rvy} = [\rvy_1, \ldots, \rvy_L]^T.
\end{split}
\end{eqnarray}
Here the complexity parameter $\lambda > 0$ controls the shrinkage of the coefficients of $g(y)$.

For kernel smoother, the processing function is (see, e.g., \cite[Chap. 6]{hastie09:book})
\begin{eqnarray}
g(y) = \frac{\sum_{l = 1}^L K_\lambda(y, \rvy_l) \rvx_l}{\sum_{l = 1}^L K_\lambda(y, \rvy_l)},
\end{eqnarray}
and in this paper we use the Gaussian kernel, $K_\lambda(y, \rvy_l) = \frac{1}{\sqrt{2\pi}\lambda} e^{-\frac{\|y - \rvy_l\|^2}{2\lambda^2}}$ in which $\lambda > 0$ controls the width of kernel.

Unless stated overwise, the channel signal-to-noise ratio (SNR) is set as $20$dB, the channel coefficients are set the same as those in the numerical example in the appendix, the size of training data set is $L = 800$,\footnote{Such a number may seem absurd for wireless communications engineers. For mobile wireless channels, it certainly does not make sense to transmit a pilot sequence of that length. As remarked in the introduction, the terminology of training in our problem is different from pilot-assisted channel training. Here the purpose of training is to learn about the statistical channel model from scratch. Section \ref{subsec:channel-state} contains some further discussion about information transmission over channels with state.} and the CV number is $Q = 5$. Each case is obtained with $10^4$ Monte Carlo simulations.

Tables II and III show $(\mathcal{P}_\mathrm{oe}, \mathcal{L}_\mathrm{r})$ for the three channel models with ridge regression and kernel smoother, respectively. A general observation is that the results exhibit diverse trends. For ridge regression, moderate values of $\lambda$ achieve
a good balance between $\mathcal{P}_\mathrm{oe}$ and $\mathcal{L}_\mathrm{r}$. For kernel smoother, $\mathcal{P}_\mathrm{oe}$ prefers a large $\lambda$ while $\mathcal{L}_\mathrm{r}$ prefers a small $\lambda$, for the first two channel models, but the trend is different for the last channel model, where both $\mathcal{P}_\mathrm{oe}$ and $\mathcal{L}_\mathrm{r}$ prefer a relatively small $\lambda$. This perhaps hints a complicated interaction between the nonlinearity of channel and the nonlinearity of kernel smoother. For most of the studied channel models, by tuning parameters (e.g., $\lambda$) in the learning algorithm, it is possible to achieve $\mathcal{L}_\mathrm{r}$ below 20\% while maintaining $\mathcal{P}_\mathrm{oe}$ at a level of 5\%. Realizing that the {\sf LFIT} algorithm is purely data-driven and does not utilize any knowledge about the statistical channel model (e.g., channel coefficients, noise, quantization, usage of dithering, etc.), such relatively small performance losses are promising sign of the potential of machine learning techniques for information transmission.

\begin{table*}[h!]
\label{table:RR}
\begin{center}
\caption{Performance with ridge regression}
\begin{tabular}{l|l|l|l|l|l|l|l|l|l|l}
\hline
\multirow{3}{*}{
    \begin{tabular}{c}
        without quantization\\
        $\xi_1 = 1.002$, $\xi_2 = 0.998$
    \end{tabular}} & $\lambda$ &0   & 100   &   200  & 1200  &  5200  &  25000  & 77000\\
\cline{2-9}
~ & $\mathcal{P}_\mathrm{oe}$ (\%) & 1.27 &   1.27   & 1.30   & 1.37  &  1.57  &  3.07  &  7.03\\
\cline{2-9}
~ &  $\mathcal{L}_\mathrm{r}$ (\%)   & 18.85  & 18.82  & 18.81  & 18.82  & 19.00  & 20.10  & 23.45\\
\hline
\multirow{3}{*}{
    \begin{tabular}{c}
        one-bit quantization\\
        $\xi_1 = 1.001$, $\xi_2 = 0.98$
    \end{tabular}} & $\lambda$ &0  & 50 & 100  & 200 & 600  & 1200  & 5200  & 9500 & 17000\\
\cline{2-11}
~ & $\mathcal{P}_\mathrm{oe}$ (\%) &0.00   & 0.00  &  0.00  &  0.00  &  0.067  &  0.067  &  1.13  &  2.60 & 4.90\\
\cline{2-11}
~ &  $\mathcal{L}_\mathrm{r}$ (\%) &18.38 &  18.30  & 18.29  & 18.31  & 18.48  & 18.70 &  20.32  & 22.64 &  27.56\\
\hline
\multirow{3}{*}{
    \begin{tabular}{c}
        dithered one-bit quantization\\
        $\xi_1 = 1.003$, $\xi_2 = 0.987$
    \end{tabular}} & $\lambda$ &0  & 50 & 100  & 200 & 600  & 1200  & 5200  & 9500 \\
\cline{2-10}
~ & $\mathcal{P}_\mathrm{oe}$ (\%) &17.47   & 1.83  &  0.40  &  0.07  &  0.07  &  0.17  &  1.67  &  4.00 \\
\cline{2-10}
~ &  $\mathcal{L}_\mathrm{r}$ (\%) &13.64 &  14.12  & 15.03  & 16.58  & 19.66  & 21.47 &  26.81  & 32.39 \\
\hline
    \end{tabular}
    \end{center}
    \end{table*}

\begin{table*}[h!]
    \label{table:KS}
    \begin{center}
    \caption{Performance with kernel smoother}
    \begin{tabular}{l|l|l|l|l|l|l|l|l|l|l|l|l}
    \hline
    \multirow{3}{*}{
        \begin{tabular}{c}
            without quantization\\
            $\xi_1 = 1.0015$, $\xi_2 = 0.9985$
        \end{tabular}} & $\lambda$ &2.80 &5.20  &8.00  &9.50 &11.00 &12.50 &14.00 &15.50   &17.00     &20.00\\
        \cline{2-12}
        ~ & $\mathcal{P}_\mathrm{oe}$ (\%) &11.85 &11.52  &10.75  &9.91  &8.67  &7.57  &5.89  &4.08  &3.01  &2.71\\
        \cline{2-12}
        ~ &  $\mathcal{L}_\mathrm{r}$ (\%)  &10.70 &10.78 &11.36 &11.96 &12.86 &14.18 &15.84 &17.91 &20.37  &26.06\\
        \hline
        \multirow{3}{*}{
            \begin{tabular}{c}
                one-bit quantization\\
                $\xi_1 = 1.01$, $\xi_2 = 0.99$
            \end{tabular}} & $\lambda$ &1.00  &2.20 &2.80 &3.60 &4.00 &4.40    &5.20  &5.60  &6.00  &6.40  &6.80\\
        \cline{2-13}
        ~ & $\mathcal{P}_\mathrm{oe}$ (\%) &0.00 &0.00  &0.00  &0.10  &0.10  &0.15 &0.40 &0.90  &1.20 &1.70 &2.20\\
        \cline{2-13}
        ~ &  $\mathcal{L}_\mathrm{r}$ (\%) &24.32  &24.79  &22.89 &22.82 &23.45 &24.42 &27.26 &29.01 &31.00  &33.15   &35.44\\
        \hline
        \multirow{3}{*}{
            \begin{tabular}{c}
                dithered one-bit quantization\\
                $\xi_1 = 1.01$, $\xi_2 = 0.99$
            \end{tabular}} & $\lambda$ &0.40   &1.00  &1.40  &2.80 &3.60 &4.00 &4.40 &5.20  &5.60 &6.00  &6.40 \\
        \cline{2-13}
        ~ & $\mathcal{P}_\mathrm{oe}$ (\%) &5.00   &5.75 &5.85   &5.85  &5.60  &5.35  &5.95 &7.75  &8.30  &9.30 &10.25 \\
        \cline{2-13}
        ~ &  $\mathcal{L}_\mathrm{r}$ (\%) &7.31   &7.83 &10.31    &17.37  &21.93 &24.68 &27.77 &34.50  &38.22  &41.94 &45.65 \\
        \hline
            \end{tabular}
            \end{center}
            \end{table*}

We further examine the $(\mathcal{P}_\mathrm{oe}$, $\mathcal{L}_\mathrm{r})$ performance when we change the configuration. Table IV compares the effects of using different scaling factors $(\xi_1, \xi_2)$ in the {\sf LFIT} algorithm, for dithered one-bit quantization, with ridge regression. We observe that by letting $(\xi_1, \xi_2)$ deviate from one, $\mathcal{P}_\mathrm{oe}$ decreases, while $\mathcal{L}_\mathrm{r}$ increases. This is because larger biases in estimating the expectations in (\ref{eqn:gmi-gT-optJ}) reduce the chance of over-estimation, as argud in the proof of Proposition \ref{prop:Poe-guarantee}, but meanwhile take a toll on the rate $R_\mathcal{T}$ calculated according to (\ref{eqn:R_T-scaling}).

\begin{table*}[h!]
    \label{table:scaling}
    \begin{center}
    \caption{Performance with different scaling factors}
    \begin{tabular}{l|l|l|l|l|l|l|l|l|l}
    \hline
    \multirow{3}{*}{
        \begin{tabular}{c}
            $\xi_1 = 1.001$, $\xi_2 = 0.99$
        \end{tabular}} & $\lambda$ &0 &50  &100  &200 &600 &1200 &5200 &9500\\
        \cline{2-10}
        ~ & $\mathcal{P}_\mathrm{oe}$ (\%) &29.33 &6.33  &2.00  &0.77  &0.57  &0.70  &3.13  &5.50\\
        \cline{2-10}
        ~ &  $\mathcal{L}_\mathrm{r}$ (\%)  &11.69 &12.14 &13.06 &14.66 &17.89 &19.79 &25.30 &31.03\\
        \hline
        \multirow{3}{*}{
            \begin{tabular}{c}
                $\xi_1 = 1.003$, $\xi_2 = 0.987$
            \end{tabular}} & $\lambda$ &0 &50  &100  &200 &600 &1200 &5200 &9500\\
        \cline{2-10}
        ~ & $\mathcal{P}_\mathrm{oe}$ (\%) &17.47 &1.83  &0.40  &0.07  &0.07  &0.17 &1.67 &4.00 \\
        \cline{2-10}
        ~ &  $\mathcal{L}_\mathrm{r}$ (\%) &13.64  &14.12  &15.03 &16.58 &19.66 &21.47 &26.81 &32.39\\
        \hline
        \multirow{3}{*}{
            \begin{tabular}{c}
                $\xi_1 = 1.005$, $\xi_2 = 0.985$
            \end{tabular}} & $\lambda$ &0 &50  &100  &200 &600 &1200 &5200 &9500\\
        \cline{2-10}
        ~ & $\mathcal{P}_\mathrm{oe}$ (\%) &11.37   &0.80 &0.10   &0.00  &0.00  &0.03  &1.17 &2.87 \\
        \cline{2-10}
        ~ &  $\mathcal{L}_\mathrm{r}$ (\%) &15.08   &15.57 &16.47    &17.96  &20.94 &22.69 &27.88 &33.40\\
        \hline
            \end{tabular}
            \end{center}
            \end{table*}

\begin{figure}
\centering
    \begin{minipage}{8cm}
    \includegraphics[width=3.1in]{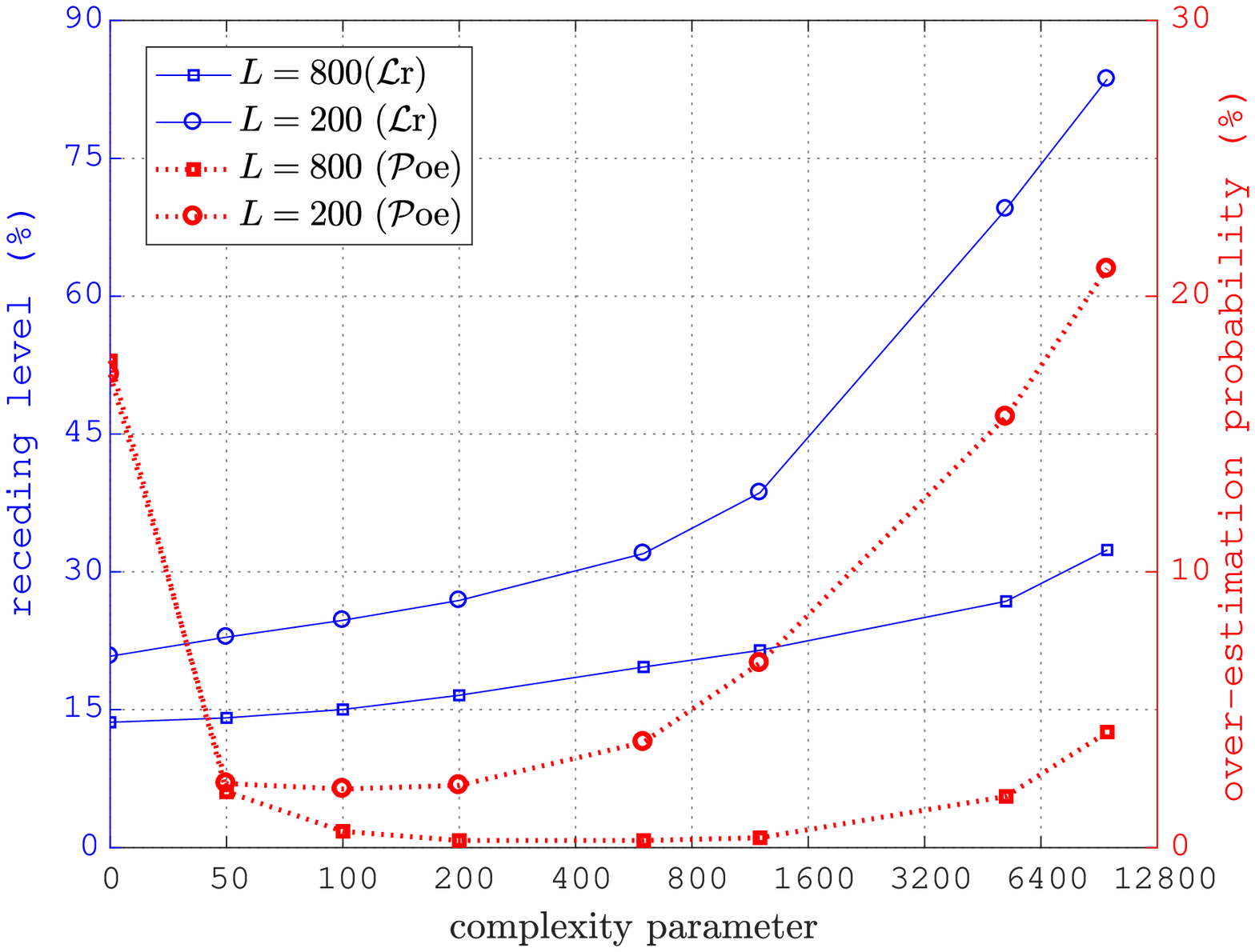}
    \caption{Comparison for different training data set sizes,\newline for dithered one-bit quantization, ridge regression.}
    \label{fig:Dither-RR-Tsize}
    \end{minipage}
    \begin{minipage}{8cm}
    \includegraphics[width=3.1in]{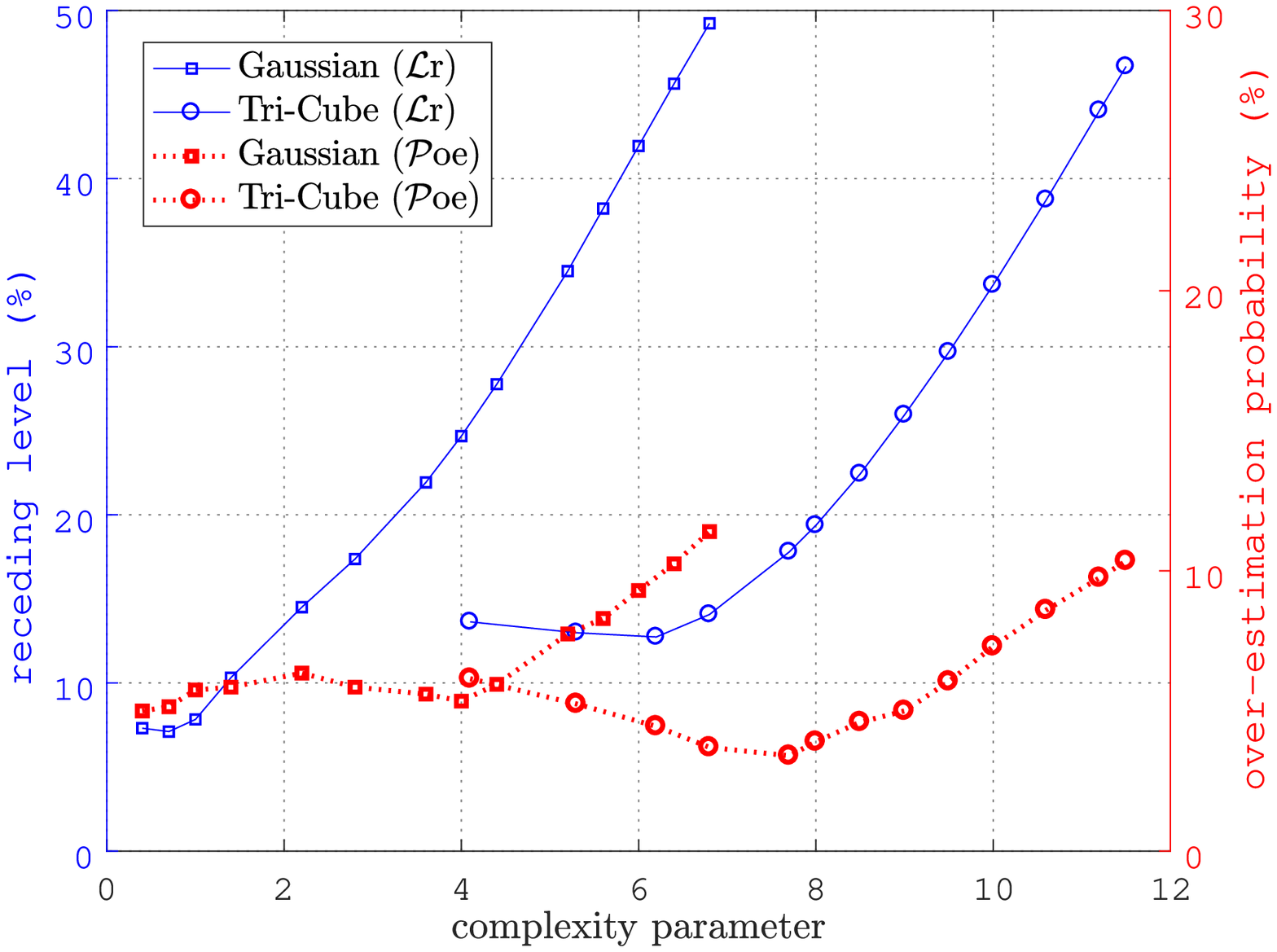}
    \caption{Comparison for different kernels, for dithered\newline one-bit quantization, kernel smoother.}
    \label{fig:Dither-KS-kernel}
    \end{minipage}
\end{figure}

\begin{figure}
    \centering
        \begin{minipage}{8cm}
        \includegraphics[width=3.1in]{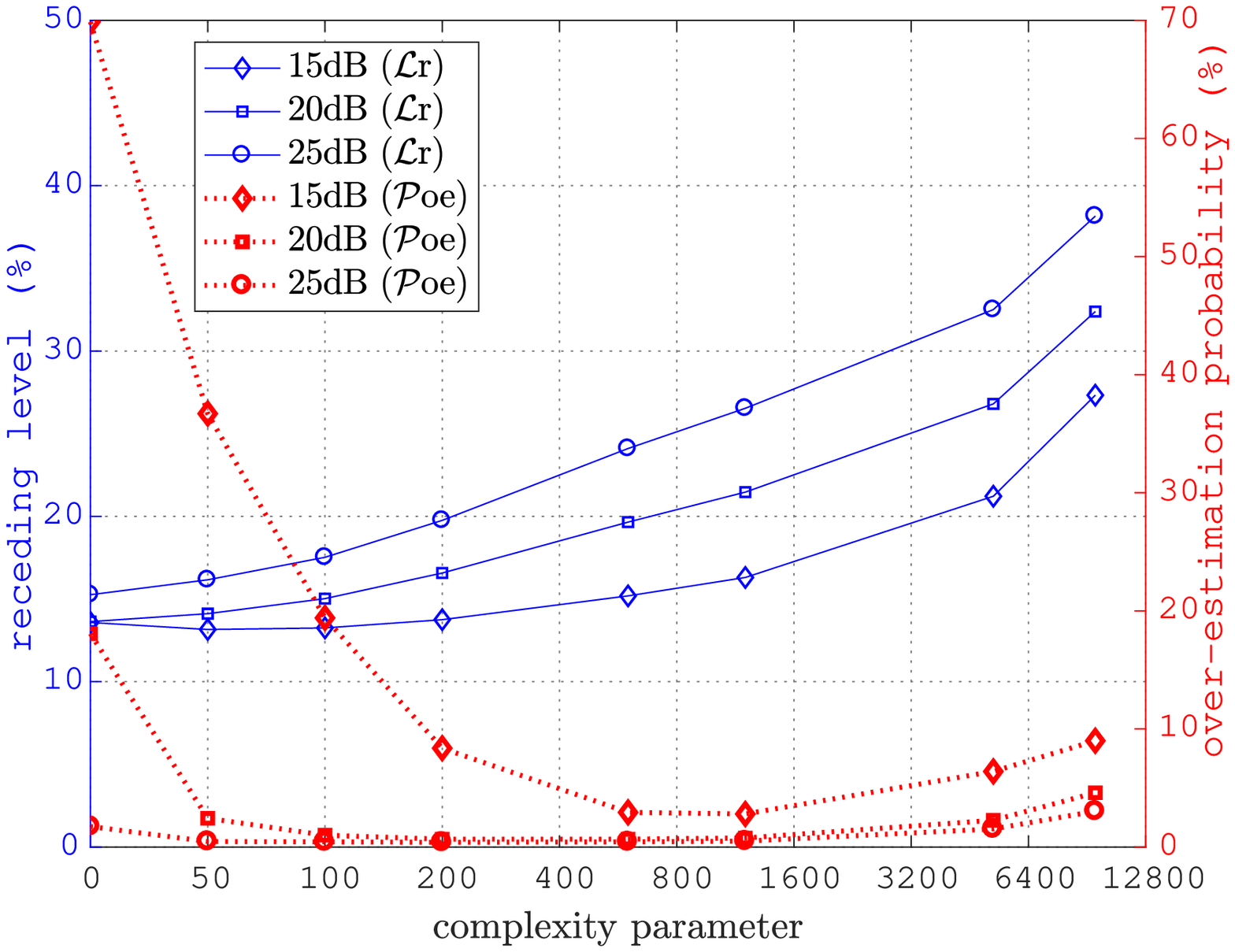}
        \caption{Comparison for different SNRs,\newline for dithered one-bit quantization, ridge regression.}
        \label{fig:Dither-RR-SNR}
        \end{minipage}
        \begin{minipage}{8cm}
        \includegraphics[width=3.1in]{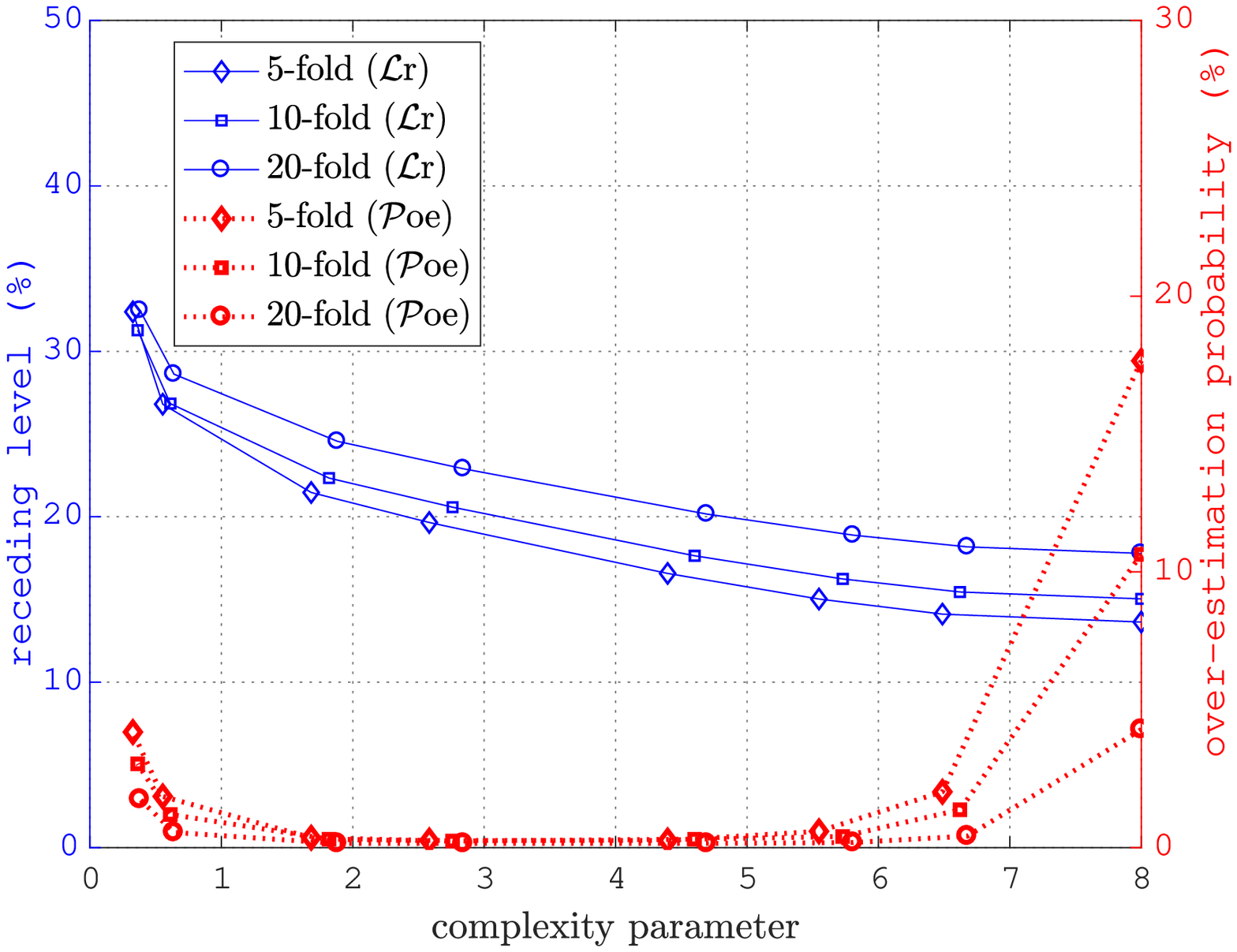}
        \caption{Comparison for different CV numbers,\newline for dithered one-bit quantization, ridge regression.}
        \label{fig:Dither-RR-CV}
        \end{minipage}
\end{figure}

Figure \ref{fig:Dither-RR-Tsize} compares the performance with different training data set sizes. By reducing $L$ from 800 to 200, we clearly observe that both $\mathcal{P}_\mathrm{oe}$ and $\mathcal{L}_\mathrm{r}$ are increased, while their trends with $\lambda$ remain largely unchanged. Figure \ref{fig:Dither-KS-kernel} compares the performance with different kernels. We consider the Gaussian kernel and the tricube kernel \cite[Chap. 6, Sec. 1, (6.6)]{hastie09:book}. We observe that although $(\mathcal{P}_\mathrm{oe}, \mathcal{L}_\mathrm{r})$ exhibit generally similar trends with $\lambda$ for the two kernels, their favorable choices of $\lambda$ (namely, width of kernel) are quite different. Figure \ref{fig:Dither-RR-SNR} compares the performance with different SNRs. As SNR increases, $\mathcal{P}_\mathrm{oe}$ decreases, but $\mathcal{L}_\mathrm{r}$ increases. Finally, Figure \ref{fig:Dither-RR-CV} compares the performance with different numbers of CV, i.e., $Q$. We observe that the choice of $Q$ has a relatively small impact on performance, especially in terms of $\mathcal{P}_\mathrm{oe}$.

\section{Discussion on Extensions}
\label{sec:extensions}

Our results about the basic channel model considered in the past two sections may be extended in several directions, several of which are briefly discussed in this section.

\subsection{Channels with Memory}
\label{subsec:channel-memory}

In many applications, the channel may not be memoryless. Suppose that during the transmission of a codeword, the channel output is a segment drawn from a stationary and ergodic stochastic process. The idea is to group consecutive input/output symbols as ``super symbols'', and consider coding over such super symbols. We thus slightly modify conditions (1) and (2) in Section \ref{sec:memoryless} as follows:

(1a) Under average power constraint $P$, super symbol length $S$, codeword length $N$, and code rate $R$ (nats/channel use), the encoder uses an i.i.d. Gaussian codebook, each of which is an independent $NS$-dimensional $\mathcal{N}(\mathbf{0}, P \mathbf{I}_{NS})$-distributed random vector.

(2a) Given a length-$NS$ channel output block $\underline{y} = [\underline{y}_1, \ldots, \underline{y}_N]$, where $\underline{y}_n$ is the $n$-th length-$S$ super symbol, the decoder, with a prescribed processing function $g: \mathcal{Y}^S \mapsto \mathbb{R}^S$ and a scaling parameter $a$, decides the transmitted message as
\begin{eqnarray}
\hat{m} = \mathrm{argmin}_{m \in \{1, 2, \ldots, M\}} D(m),\nonumber\\
\mbox{where}\; D(m) = \sum_{n = 1}^N \left\|g(\underline{y}_n) - a \underline{x}_n(m) \right\|^2,\label{eqn:nn-metric-memory}
\end{eqnarray}
and $\underline{x}_n(m)$ is the length-$S$ super symbol in the codeword corresponding to message $m$.

We have the following result.
\begin{prop}
\label{prop:gmi-memory}
For an information transmission model under conditions (1a) and (2a), assuming that the normalized MMSE of estimating an input super symbol $\underline{\rvx}$ upon observing its corresponding output super symbol $\underline{\rvy}$ has a limit as $S \rightarrow \infty$, i.e., $\mathsf{mmse} = \lim_{S \rightarrow \infty} \frac{1}{S} \mathbf{E}\left[\left\|\underline{\rvx} - \mathbf{E}\left[\underline{\rvx}|\underline{\rvy}\right]\right\|^2\right]$ existing, the information rate
\begin{eqnarray}
I_{\mathrm{GMI}, \mathrm{MMSE}} = \frac{1}{2} \log \frac{P}{\mathsf{mmse}}
\end{eqnarray}
is achievable, when $g$ is the MMSE smoother $g(\underline{y}) = \mathbf{E}[\underline{\rvx}|\underline{y}]$.
\end{prop}
{\it Proof:} The proof is essentially a combination and extension of the proofs of Propositions \ref{prop:generic-g} and \ref{prop:nonlinear}. Here we give its outline. Fix $S$ and $g$. Due to symmetry in the codebook, assume that $m = 1$ is the transmitted message. So due to the strong law of large numbers, $\lim_{N \rightarrow \infty} D(1) = \mathbf{E}\left[\left\|g(\underline{\rvy}) - a \underline{\rvx}\right\|^2\right]$, a.s., and the GMI is then given by
\begin{equation}
I_{\mathrm{GMI}, g} \!\!=\!\! \sup_{\theta < 0, a} \Big\{\theta \mathbf{E}\left[\left\|g(\underline{\rvy}) - a \underline{\rvx}\right\|^2\right] - \Lambda(\theta)\Big\},
\end{equation}
\begin{equation}
\begin{split}
\Lambda(\theta) &\!\!=\!\! \lim_{N \rightarrow \infty} \frac{1}{N} \Lambda_N (N \theta), \\
\Lambda_N(N \theta) &\!\!=\!\! \log \mathbf{E} \left[e^{N \theta D(m)} \Big | \underline{\rvy}_1, \ldots, \underline{\rvy}_N \right], \; \forall m \neq 1.
\end{split}
\end{equation}
Following a similar approach as in the proof of Proposition \ref{prop:generic-g}, we can derive
\begin{eqnarray}
\Lambda(\theta) = \frac{\theta \mathbf{E}\left[\|g(\underline{\rvy})\|^2\right]}{1 - 2\theta a^2 P} - \frac{S}{2} \log (1 - 2\theta a^2 P), \; \mbox{a.s.}
\end{eqnarray}
Then we can solve for $I_{\mathrm{GMI}, g}$ as a vector extension of Proposition \ref{prop:generic-g}, with an essentially identical solution procedure. The optimal $a$ is $a = \frac{\mathbf{E}\left[\underline{\rvx}^T g(\underline{\rvy})\right]}{S P}$, and the GMI is
$I_{\mathrm{GMI}, g} = \frac{1}{2} \log \frac{1}{1 - \Delta_{S, g}}$, $\Delta_{S, g} = \frac{\left(\mathbf{E}\left[\underline{\rvx}^T g(\underline{\rvy})\right]\right)^2}{SP \mathbf{E}\left[\|g(\underline{\rvy})\|^2\right]}$. By extending the definition of correlation ratio to vector random variables, and by mimicking the proof of Lemma \ref{lem:renyi-correlation-coefficient}, we can show that $\max_g \Delta_{S, g} = \frac{\mbox{tr}\left[\mbox{cov} \mathbf{E}[\underline{\rvx}| \underline{\rvy}]\right]}{S P}$, attained by choosing $g(\underline{y}) = \mathbf{E}[\underline{\rvx} | \underline{y}]$. This leads to
$I_{\mathrm{GMI}, g} = \frac{1}{2} \log \frac{P}{\mathsf{mmse}_S}$,
where $\mathsf{mmse}_S = (1/S) \mathbf{E}\left[\|\underline{\rvx} - \mathbf{E}[\underline{\rvx} | \underline{\rvy}]\|^2\right]$ is the normalized MMSE of estimating $\underline{\rvx}$ upon observing $\underline{\rvy}$, which by our assumption has a limit $\mathsf{mmse}$ as $S \rightarrow \infty$. This thus completes the proof. {\bf Q.E.D.}

In view of Proposition \ref{prop:gmi-memory}, we conclude that for a channel with memory, we can also formulate a learning-based problem formulation and develop algorithms following the line of Section \ref{sec:learning}, by working with super symbols. A practical challenge, of course, is the curse of dimensionality, as the super symbol length $S$ grows large.

\subsection{Channels with General Inputs and General Decoding Metrics}
\label{subsec:general-input}

The conditions (1) and (2) in Section \ref{sec:memoryless} are restrictive in terms of input distribution and decoding metric. Mismatched decoding is, in principle, capable of handling channels with general inputs and general decoding metrics \cite{ganti00:it}. Suppose that the i.i.d. codebook is $p(x)$-distributed, $x \in \mathcal{X}$, and that the decoding metric in (\ref{eqn:nn-metric}) is replaced by $\sum_{n = 1}^N d(x_n(m), y_n)$ for a generic input-output distance metric $d$. The corresponding GMI can be expressed as \cite{ganti00:it}
\begin{eqnarray}
\label{eqn:gmi-general-input-metric}
I_{\mathrm{GMI}, g, d} = \sup_{\theta \geq 0} \mathbf{E} \left[\log \frac{e^{-\theta d(\rvx, \rvy)}}{\mathbf{E} \left[ e^{-\theta d(\rvx, \rvy)}\big |\rvy\right]}\right].
\end{eqnarray}
Here the expectation in the denominator is with respect to $p(x)$ only.

Although a closed-form expression of (\ref{eqn:gmi-general-input-metric}) may not be available, we can formulate the problem of optimizing the distance metric $d$ (possibly subject to certain structural constraint) to maximize (\ref{eqn:gmi-general-input-metric}). Furthermore, without utilizing the statistical channel model, we can formulate the problem of learning a good choice of $d$ based upon training data, and investigate the behavior of over-estimation probability and receding level, as done in Section \ref{sec:learning}.

\subsection{Channels with State and Channel Estimation}
\label{subsec:channel-state}

Consider a channel with random state $\rvs$ which is independent of channel input and changes in an i.i.d. fashion across consecutive length-$B$ blocks. This is a commonly adopted model for describing time-varying wireless fading channels, where the fading coefficient corresponds to channel state. We may view each length-$B$ block as a super symbol and write the channel input-output conditional probability law as $p(\underline{y}|\underline{x}) = \sum_{s \in \mathcal{S}} p(s) \prod_{i = 1}^B p(y_i|x_i, s)$, $(\underline{x}, \underline{y}, s) \in \mathcal{X}^B \times \mathcal{Y}^B \times \mathcal{S}$ (see, e.g., \cite[Chap. 7, Sec. 4]{elgamal:book}). This way, we return to a memoryless channel without state. Under scenario (A) introduced in Section \ref{subsec:formulation}, without any knowledge about the statistical channel model, we let the encoder transmit training inputs across a large number of super symbols (i.e., length-$B$ blocks), so that a learning phase is accomplished at the decoder to characterize the super symbol level channel behavior, following the general discussion in Section \ref{subsec:general-input}. This is a non-coherent approach \cite{hochwald00:it} \cite{zheng02:it}, and it is reasonable, because we do not even have any knowledge about the statistical channel model, let alone the realization of channel states.

Under scenario (B), we have knowledge about $p(y|x, s)$ as well as $p(s)$, and we may adopt some alternative approaches, combining ideas from learning and pilot-assisted channel state estimation. One such approach is outlined as follows. Similar to our description in Section \ref{subsec:formulation}, the decoder simulates the channel to generate the training data set. But here due to the existence of channel state, the generated training data set is $\mathcal{T} = \{(\rvs_1, \rvx_1, \rvy_1), \ldots, (\rvs_K, \rvx_K, \rvy_K)\}$, where $(\rvs_k, \rvx_k, \rvy_k)$ is distributed according to $p(s) p(x) p(y|x, s)$; that is, each time we sample an input $\rvx$ according to $p(x)$, a state $\rvs$ according to $p(s)$, and generate an output $\rvy$ according to $p(y|x, s)$. Based upon $\mathcal{T}$, we learn two functions: a state estimator $h_\mathcal{T}$ and a distance metric $d_\mathcal{T}$. The state estimator $h_\mathcal{T}: \mathcal{X} \times \mathcal{Y} \mapsto \mathcal{S}$ estimates the state based upon an observed input-output pair, and the distance metric $d_\mathcal{T}: \mathcal{X} \times \mathcal{Y} \times \mathcal{S} \mapsto \mathbb{R}$ depends upon both the channel input-output pair and the estimated state.

During the information transmission phase, in a length-$B$ block, the encoder spends one channel use to transmit a pilot symbol $\rvx_\mathrm{p}$ through the channel to the decoder to produce $\rvy_\mathrm{p}$, and the decoder uses the state estimator $h_\mathcal{T}$ to estimate the state in that block, denoted by $\hat{\rvs} = h_\mathcal{T}(\rvx_\mathrm{p}, \rvy_\mathrm{p})$.

Under i.i.d. $p(x)$ codebook ensemble and learnt distance metric $d_\mathcal{T}$ for decoding, the corresponding GMI can be derived, by extending (\ref{eqn:gmi-general-input-metric}), as
\begin{eqnarray}
\label{eqn:gmi-general-input-metric-state}
I_{\mathrm{GMI}, h, g, d} = \frac{B - 1}{B}\sup_{\theta \geq 0} \mathbf{E} \left[\log \frac{e^{-\theta d_\mathcal{T}(\rvx, \rvy, \hat{\rvs})}}{\mathbf{E} \left[ e^{-\theta d_\mathcal{T}(\rvx, \rvy, \hat{\rvs})}\big |\rvy\right]}\right].
\end{eqnarray}
Here the expectation in the denominator is with respect to $p(x)$ only. The $(B - 1)/B$ factor is due to the transmission of pilot symbol.

\section{Conclusion}
\label{sec:conclusion}

In this paper, we considered the problem of learning for transmitting information over a channel without utilizing its statistical model. Our approach is rudimentary: the learning phase and the information transmission phase are separated; in the information transmission phase, the encoder and the decoder have prescribed structure, namely i.i.d. Gaussian codebook ensemble and nearest neighbor decoding rule; in the learning phase, we only learn the channel output processing function and the scaling parameter for implementing the decoder, and choose a code rate. As discussed in the introduction, a full-blown learning based information transmission system should be able to adaptively learn a channel, and to adjust its codebook as well as decoding metrics accordingly, in a dynamic fashion, with the ultimate goal of approaching channel capacity.

Nevertheless, even within the scope of our problem formulation, there is still an important issue which we have yet to touch in this paper. The $\mathsf{LFIT}$ algorithm provides a design for each given parameter $\lambda$ of the processing function, but does not recommend the optimal choice of $\lambda$. Because of this, our numerical experiments in Section \ref{subsec:learning-case-study} are ``offline'' in the sense that $\mathcal{P}_\mathrm{oe}$ and $\mathcal{L}_\mathrm{r}$ can only be evaluated when the statistical channel model is given. To resolve this difficulty, it is desirable to develop methods (e.g., via bootstrap \cite{efron98:book}) to estimate $\mathcal{P}_\mathrm{oe}$ and $\mathcal{L}_\mathrm{r}$ based upon a training data set $\mathcal{T}$ only.

\section*{Appendix: Case Studies}

We consider three kinds of channel models.

\subsubsection{SIMO Channel without Quantization}

The channel model is
\begin{eqnarray}\label{eqn:simo-channel}
\rvy = \mathbf{h} \rvx + \rvz.
\end{eqnarray}
In the real case, $\rvx \sim \mathcal{N}(0, P)$, $\rvz \sim \mathcal{N}(0, \sigma^2 \mathbf{I}_p)$, and $\mathbf{h}$ is a $p$-dimensional real vector representing channel coefficients.\footnote{This channel model can also describe the case where a scalar symbol is weighted by several transmit antennas (i.e., beamforming) and then transmitted through a multiple-input-multiple-output (MIMO) channel. More general MIMO cases can be investigated following the discussion in Section \ref{subsec:general-input}.} Applying Proposition \ref{prop:linear}, we have $\Delta_\mathrm{LMMSE} = P \mathbf{h}^T \left(P \mathbf{h} \mathbf{h}^T + \sigma^2 \mathbf{I}_p\right)^{-1} \mathbf{h}$. An exercise of Woodbury matrix identity yields the corresponding GMI as
\begin{eqnarray}
\begin{split}
I_{\mathrm{GMI}, \mathrm{LMMSE}} &= \frac{1}{2} \log \frac{1}{1 - \Delta_\mathrm{LMMSE}} \\
&= \frac{1}{2} \log \left(1 + \frac{P}{\sigma^2} \|\mathbf{h}\|^2\right),
\end{split}
\end{eqnarray}
which is exactly the capacity of the channel. To understand this result, note that here the nearest neighbor decoding rule with LMMSE output estimation is matched to the channel (i.e., maximum-likelihood).

As already done in \cite[App. C]{zhang12:tcom}, our analysis in Section \ref{sec:memoryless} can be directly extended to the complex case, by changing the input distribution in condition (1) to circularly symmetric complex Gaussian distribution $\mathcal{CN}(0, P)$, and changing the distance metric in condition (2) to $|g(y_n) - a x_n(m)|^2$. Proposition \ref{prop:generic-g} still applies, with $I_{\mathrm{GMI}, g} = \log \frac{1}{1 - \Delta_g}$, $\Delta_g = \frac{\left|\mathbf{E}\left[\rvx g(\rvy)\right]\right|^2}{P \mathbf{E}\left[|g(\rvy)|^2\right]}$. Propositions \ref{prop:linear} and \ref{prop:nonlinear} also hold, by extending the definitions of involved operations to complex numbers.

For the channel model (\ref{eqn:simo-channel}), in the complex case, $\rvx \sim \mathcal{CN}(0, P)$, $\rvz \sim \mathcal{CN}(0, \sigma^2 \mathbf{I}_p)$, and $\mathbf{h}$ is $p$-dimensional complex-valued. The GMI under the LMMSE estimator again coincides with the channel capacity, i.e.,
$I_{\mathrm{GMI}, \mathrm{LMMSE}} = \log \left(1 + \frac{P}{\sigma^2} \|\mathbf{h}\|^2\right)$.

\subsubsection{SIMO Channel with One-bit Quantization}

With a one-bit quantizer at each of the output component, the channel model in (\ref{eqn:simo-channel}) becomes
\begin{eqnarray}
\rvy_i = \mathcal{Q}_\mathrm{r}(h_i \rvx + \rvz_i), \quad i = 1, \ldots, p, \;\mbox{for the real case};\\
\rvy_i = \mathcal{Q}_\mathrm{c}(h_i \rvx + \rvz_i), \quad i = 1, \ldots, p, \;\mbox{for the complex case},
\end{eqnarray}
where $\mathcal{Q}_\mathrm{r}(x) = 1$ if $x \geq 0$ and $-1$ otherwise for $x \in \mathbb{R}$, and $\mathcal{Q}_\mathrm{c}(z) = \mathcal{Q}_\mathrm{r}(z_\mathrm{r}) + \jmath \mathcal{Q}_\mathrm{r}(z_\mathrm{c})$ for $z = z_\mathrm{r} + \jmath z_\mathrm{c} \in \mathbb{C}$, $\jmath = \sqrt{-1}$.

For the real case, the MMSE estimator can be derived as
\begin{eqnarray}
\mathbf{E}[\rvx | \mathbf{y}] = \frac{\int_{-\infty}^\infty u f(u) \prod_{i = 1}^p F(y_i h_i u) \mathrm{d}u}{\int_{-\infty}^\infty f(u) \prod_{i = 1}^p F(y_i h_i u) \mathrm{d}u},
\end{eqnarray}
where $f(u)$ is the probability density function of $\mathcal{N}(0, P)$, and $F(u)$ is the cumulative distribution function of $\mathcal{N}(0, 1)$. The variance of $\mathbf{E}[\rvx | \rvy]$ is
\begin{eqnarray}
\mathrm{var} \mathbf{E}[\rvx | \rvy] = \!\!\!\!\!\!\sum_{\mathbf{y} \in \{-1, +1\}^p}\!\!\! \frac{\left(\int_{-\infty}^\infty u f(u) \prod_{i = 1}^p F(y_i h_i u) \mathrm{d}u\right)^2}{\int_{-\infty}^\infty f(u) \prod_{i = 1}^p F(y_i h_i u) \mathrm{d}u}.
\end{eqnarray}
For the complex case, the MMSE estimator and its variance can also be derived in integral form (omitted for brevity). From these we can evaluate the GMI according to Proposition \ref{prop:nonlinear}, numerically.

\subsubsection{SIMO Channel with Dithered One-bit Quantization}

A useful idea in quantization techniques is dithering, i.e., adding prescribed (possibly random) biases at the input of quantizer \cite{gray98:it}. It turns out that dithering is also beneficial in our model. Intuitively, dithering is capable of ``randomizing'' the residual quantization error, thus preserving diversity among the output components. The channel model in (\ref{eqn:simo-channel}) becomes
\begin{equation}
\label{eqn:simo-channel-1bit-dither}
\rvy_i = \mathcal{Q}_\mathrm{r} (h_i \rvx + \rvz_i + b_i), \quad i = 1, \ldots, p,\;\mbox{for the real case};
\end{equation}
\begin{equation}
\rvy_i = \mathcal{Q}_\mathrm{c} (h_i \rvx + \rvz_i + b_i), \quad i = 1, \ldots, p,\;\mbox{for the complex case},
\end{equation}
where $b_i$ is a dithering bias added to the $i$-th output component. The MMSE estimator and its variance can also be derived in integral form and are omitted here for brevity. An exhaustive search of optimal $\{b_i\}_{i = 1, \ldots, q}$ is intractable, and we use a heuristic design as $b_i = \alpha \sqrt{P} h_i u_i$, where $u_i$ is the solution of $F(u) = i/(p + 1)$ and $\alpha$ is a parameter which can then be numerically optimized.

We present some numerical examples for illustrating the GMIs computed for channels with one-bit quantization, without and with dithering. For the real case, we set $p = 8$, channel coefficients $\mathbf{h} =[0.3615,$ $0.2151, 0.2205, 0.6767, 0.5014, 0.1129, 0.1763, 0.1456]^T$, and dithering parameter $\alpha = 1.34$; for the complex case, $p = 4$, $\mathbf{h} = [-0.0165 - \jmath 0.3395, -0.3793 + \jmath 0.6422, 0.4841 - \jmath 0.1068, -0.0633 - \jmath 0.2799]^T$, and $\alpha = 0.76$.\footnote{The optimal value of $\alpha$ is SNR-dependent. Here for simplicity we just optimize it for $\mathrm{SNR} = 20$dB and use the value throughout the SNR range.} Note that for both cases we have normalized the channel so that $\|\mathbf{h}\| = 1$. In the figures, the SNR is measured as $\|\mathbf{h}\|^2 P/\sigma^2$.

\begin{figure}
\centering
    \begin{minipage}{8cm}
    \includegraphics[width=3.1in]{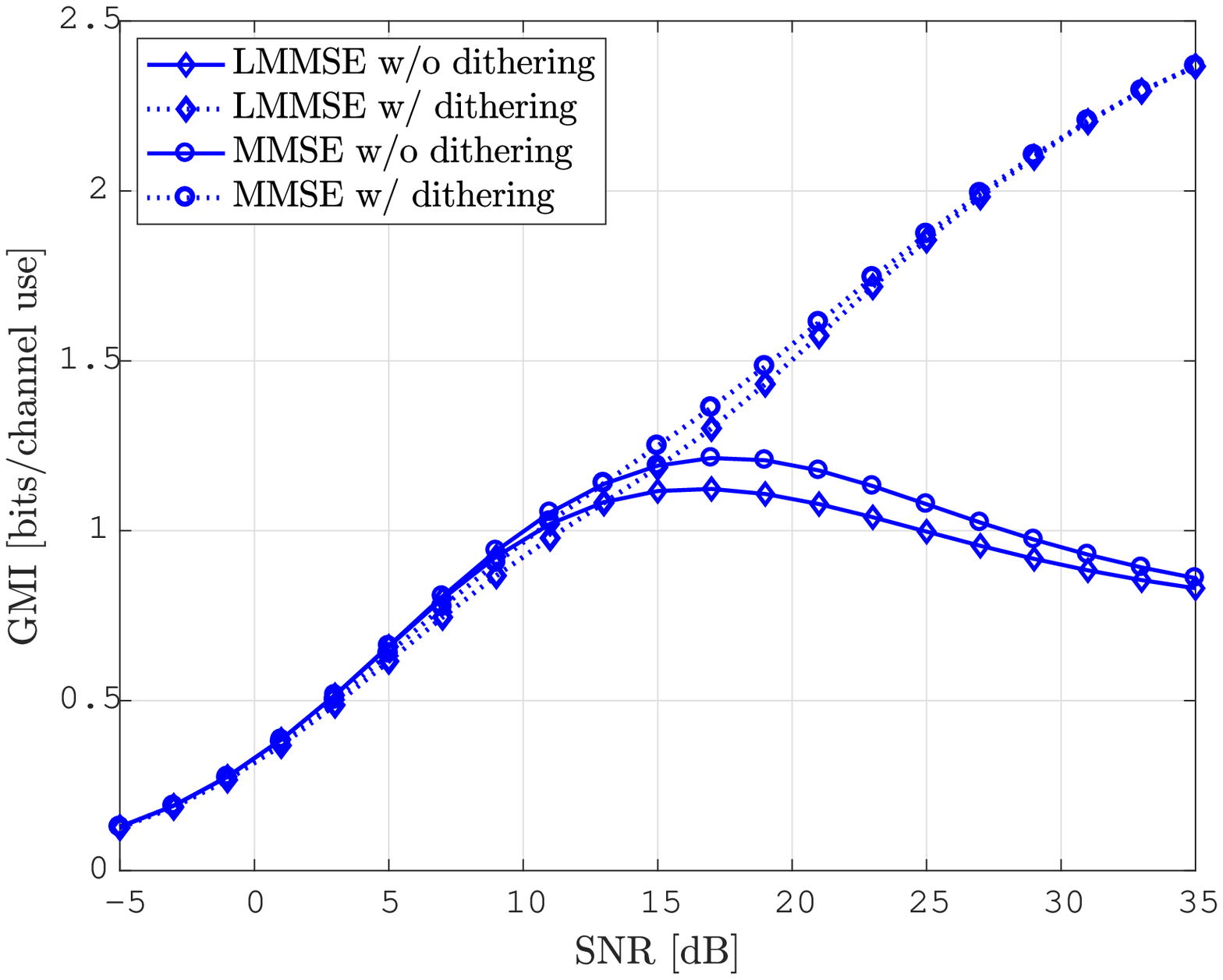}
    \caption{GMIs for the real case.}
    \label{fig:real-8}
    \end{minipage}
    \begin{minipage}{8cm}
    \includegraphics[width=3.1in]{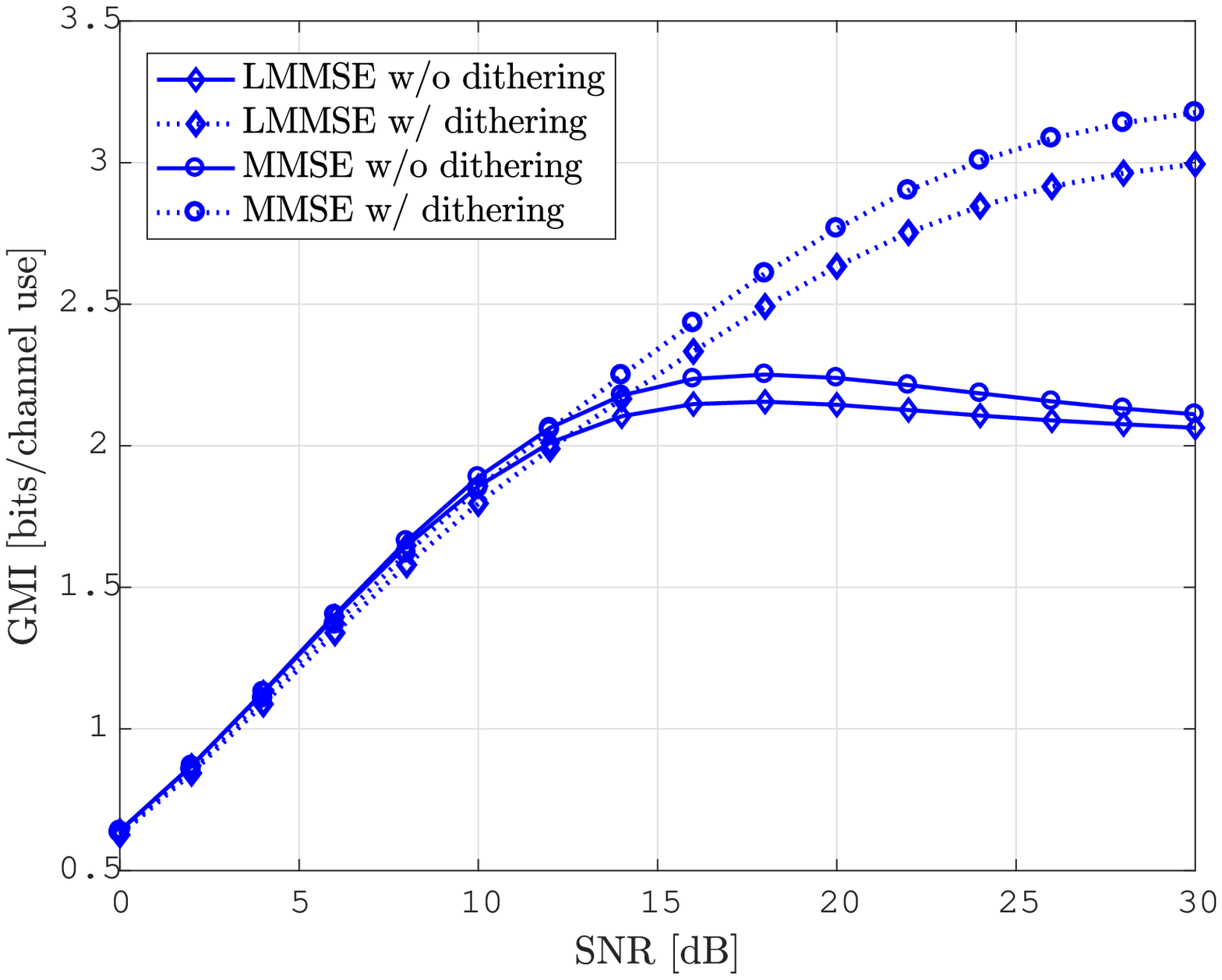}
    \caption{GMIs for the complex case.}
    \label{fig:complex-4}
    \end{minipage}
\end{figure}

Figure \ref{fig:real-8} displays the GMIs for the real case. We see that the MMSE estimator slightly outperforms the LMMSE estimator in terms of GMI, a fact already revealed in Propositions \ref{prop:linear} and \ref{prop:nonlinear}. Furthermore, dithering significantly boosts the GMIs. Without dithering, the GMI initially grows with SNR but then begins to decrease as SNR grows large. This is because the quantized channel output components eventually lose diversity as SNR grows, and thus the nearest neighbor decoding rule becomes more mismatched to the actual statistical channel model at high SNR. With dithering, the GMIs maintain their growth at least throughout the SNR range plotted. Similar observations can be made for the complex case as well, displayed in Figure \ref{fig:complex-4}.

\section*{Acknowledgement}

Stimulating discussions with Dongning Guo (Northwestern University), Tie Liu (Texas A\&M University), and H. Vincent Poor (Princeton University), and valuable comments and suggestions by the anonymous reviewers are gratefully acknowledged. Ziyu Ying has assisted in developing Proposition \ref{prop:Poe-guarantee}.



\begin{thebibliography}{1}

\bibitem{zhang16:ita} W.~Zhang, ``A remark on channels with transceiver distortion,'' in \textit{Proc. Inform. Theory and Appl. Workshop (ITA)}, Feb. 2016.

\bibitem{paninski03:nc} L.~Paninski, ``Estimation of entropy and mutual information,'' \textit{Neural Comput.}, 15(6), 1191-1253, Jun. 2003.

\bibitem{csiszar:book} I.~Csisz\'ar and J.~K\"orner, \textit{Information Theory: Coding Theorems for Discrete Memoryless Systems}, 2nd ed., Cambridge University Press, 2011.

\bibitem{lapidoth98:it} A.~Lapidoth and P.~Narayan, ``Reliable communication under channel uncertainty,'' \textit{IEEE Trans. Inform. Theory}, 44(6), 2148-2177, Oct. 1998.

\bibitem{28}
T.~J.~O'Shea, K.~Karra, and T. C.~Clancy, ``Learning to communicate: channel auto-encoders, domain specific regularizers, and attention,'' in \textit{Proc. IEEE Int. Symp. Signal Process. Inf. Technol. (ISSPIT)}, Dec. 2016.

\bibitem{dorner} S.~D\"orner, S.~Cammerer, J.~Hoydis, and S.~ten Brink, ``Deep learning based communication over the air,'' \textit{IEEE J. Sel. Topics Signal Process.}, 12(1), 132-143, Feb. 2018.

\bibitem{balevi}
E.~Balevi and J.~G.~Andrews, ``One-bit OFDM receivers via deep learning,'' arXiv:1811.00971, 2018.

\bibitem{kim18:nips} H.~Kim, Y.~Jiang, S.~Kannan, S.~Oh, and P.~Viswanath, ``Deepcode: feedback codes via deep learning,'' in \textit{Proc. Neural Information Processing Systems (NIPS)}, 2018.

\bibitem{29}
E.~Nachmani, Y.~Beery, and D.~Burshtein, ``Learning to decode linear codes using deep learning,'' in \textit{Proc. IEEE Annu. Allerton Conf. Commun., Control, Comput.}, Sep. 2016.

\bibitem{27}
W.~Lyu, Z.~Zhang, C.~Jiao, K.~Qin, and H.~Zhang, ``Performance evaluation of channel decoding with deep neural networks,'' in \textit{Proc. IEEE Int. Conf. Commun. (ICC)}, May 2018.

\bibitem{liang}
F.~Liang, C.~Shen, and F.~Wu, ``An iterative BP-CNN architecture for channel decoding,'' \textit{IEEE J. Sel. Topics Signal Process.}, 12(1), 144-159, Feb. 2018.

\bibitem{14}
H.~Ye and G. Y.~Li, ``Initial results on deep learning for joint channel equalization and decoding,'' in \textit{Proc. IEEE Veh. Tech. Conf. (VTC)}, Jun. 2017.

\bibitem{15}
W.~Xu, Z.~Zhong, Y.~Beery, X.~You, and C.~Zhang, ``Joint neural network equalizer and decoder,'' in \textit{Proc. Int. Symp. Wireless Commun. Syst. (ISWCS)}, Aug. 2018.

\bibitem{11}
A.~Caciularu and D.~Burshtein, ``Blind channel equalization using variational autoencoders,'' in \textit{Proc. IEEE Int. Conf. Commun. Workshops (ICC Workshops)}, May 2018.

\bibitem{3}
N.~Samuel, T.~Diskin, and A.~Wiesel, ``Learning to detect,'' arXiv:1805.07631, 2018.

\bibitem{1}
H.~He, C.~Wen, S.~Jin, and G. Y.~Li, ``A model-driven deep learning network for MIMO detection,'' arXiv:1809.09336, 2018.

\bibitem{26}
Q.~Huang, C.~Zhao, M.~Jiang, X.~Li, and J.~Liang, ``Cascade-Net: a new deep learning architecture for OFDM detection,'' arXiv:1812.00023, 2018.

\bibitem{4}
S.~Navabi, C.~Wang, O. Y.~Bursalioglu, and H.~Papadopoulos, ``Predicting wireless channel features using neural networks,'' in \textit{Proc. IEEE Int. Conf. Commun. (ICC)}, May 2018.

\bibitem{16}
D.~Neumann, H.~Wiese, and W.~Utschick, ``Learning the MMSE channel estimator,'' \textit{IEEE Trans. Signal Process.}, 66(11), 2905-2917, Jun. 2018.

\bibitem{17}
Y.~Wang, M.~Narasimha, and R.~Heath, ``MmWave beam prediction with situational awareness: a machine learning approach,'' in \textit{Proc. IEEE Int. Workshop Signal Process. Adv. Wireless Commun. (SPAWC)}, Jun. 2018.

\bibitem{7}
H.~He, C.~Wen, S.~Jin, and G. Y.~Li, ``Deep learning-based channel estimation for beamspace mmWave massive MIMO systems,'' \textit{IEEE Wireless Commun. Lett.}, 7(5), 852 - 855, Oct. 2018.

\bibitem{23}
R.~Deng, Z.~Jiang, S.~Zhou, S.~Cui, and Z.~Niu, ``A two-step learning and interpolation method for location-based channel database,'' arXiv:1812.01247, 2018.

\bibitem{18}
N.~Farsad and A.~Goldsmith, ``Neural network detection of data sequences in communication systems,'' \textit{IEEE Trans. Signal Process.}, 66(21), 5663-5678, Nov. 2018.

\bibitem{cover06:book} T.~M.~Cover and J.~A.~Thomas, \textit{Elements of Information Theory}, 2nd ed., Wiley-Interscience, 2006.

\bibitem{scarlett14:phd} J.~Scarlett, \textit{Reliable Communication under Mismatched Decoding}, Ph.D. dissertation, University of Cambridge, UK, 2014.

\bibitem{hui83:phd} J.~Hui, \textit{Fundamental Issues of Multiple Accessing}, Ph.D. dissertation, Massachusetts Institute of Technology, USA, 1983.

\bibitem{kaplan93:aeu} G.~Kaplan and S.~Shamai, ``Information rates and error exponents of compound channels with application to antipodal signaling in a fading environment,'' \textit{Arch. Elek. Uber.}, 47(4), 228-239, 1993.

\bibitem{merhav94:it} N.~Merhav, G.~Kaplan, A.~Lapidoth, and S.~Shamai, ``On information rates for mismatched decoders,'' \textit{IEEE Trans. Inform. Theory}, 40(6), 1953-1967, Nov. 1994.

\bibitem{ganti00:it} A.~Ganti, A.~Lapidoth, and E.~Telatar, ``Mismatched decoding revisited: general alphabets, channels with memory, and the wide-band limit,'' \textit{IEEE Trans. Inform. Theory}, 46(7), 2315-2328, Nov. 2000.

\bibitem{somekh15:it} A.~Somekh-Baruch, ``A general formula for the mismatch capacity,'' \textit{IEEE Trans. Inform. Theory}, 61(9), 4554-4568, Sep. 2015.

\bibitem{zhang12:tcom} W.~Zhang, ``A general framework for transmission with transceiver distortion and some applications,'' \textit{IEEE Trans. Commun.}, 60(2), 384-399, Feb. 2012.

\bibitem{lapidoth02:it} A.~Lapidoth and S.~Shamai (Shitz), ``Fading channels: how perfect need `perfect side information' be?'' \textit{IEEE Trans. Inform. Theory}, 48(5), 1118-1134, May 2002.

\bibitem{bjornson14:it} E.~Bj\"ornson, J.~Hoydis, M.~Kountouris, and M.~Debbah, ``Massive MIMO systems with non-ideal hardware: energy efficiency, estimation, and capacity limits,'' \textit{IEEE Trans. Inform. Theory}, 60(11), 7112-7139, Nov. 2014.

\bibitem{horn:book}
R.~A.~Horn and C.~A.~Johnson, \textit{Matrix Analysis}, Cambridge University Press, 1985.

\bibitem{poor94:book} H.~V.~Poor, \textit{An Introduction to Signal Detection and Estimation}, 2nd ed., Springer, 1994.

\bibitem{liang16:jsac} N.~Liang and W.~Zhang, ``Mixed-ADC massive MIMO,'' \textit{IEEE J. Select. Areas Commun.} 34(4), 983-997, Apr. 2016.

\bibitem{li17:vtc} B.~Li, N.~Liang, and W.~Zhang, ``On transmission model for massive MIMO under low-resolution output quantization,'' in \textit{Proc. IEEE Veh. Tech. Conf. (VTC)}, Jun. 2017.

\bibitem{bussgang52:rle} J.~J.~Bussgang, ``Crosscorrelation functions of amplitude-distorted Gaussian signals,'' TR 216, RLE, Massachusetts Institute of Technology, Cambridge, MA, USA, Mar. 1952.

\bibitem{rowe82:bstj} H.~E.~Rowe, ``Memoryless nonlinearities with Gaussian inputs: elementary results,'' \textit{Bell Syst. Tech. J.}, 61(7), 1519-1525, Sep. 1982.

\bibitem{ochiai02:tcom} H.~Ochiai and H.~Imai, ``Performance analysis of deliberately clipped OFDM signals,'' \textit{IEEE Trans. Commun.}, 50(1), 89-101, Jan. 2002.

\bibitem{orhan15:ita} O.~Orhan, E.~Erkip, and S.~Rangan, ``Low power analog-to-digital conversion in millimeter wave systems: impact of resolution and bandwidth on performance,'' in \textit{Proc. Information Theory and Applications (ITA) Workshop}, 2015.

\bibitem{renyi59:amash} A.~R\'{e}nyi, ``New version of the probabilistic generalization of the large sieve,'' \textit{Acta Math. Acad. Sci. Hung.}, 10, 218-226, 1959.

\bibitem{hastie09:book} T.~Hastie, R.~Tibshirani, and J.~Friedman, \textit{The Elements of Statistical Learning}, 2nd ed., Springer, 2009.

\bibitem{billingsley:book} P.~Billingsley, \textit{Probability and Measure}, 3rd ed., John Wiley and Sons, 1995.

\bibitem{boucheron:book} S.~Boucheron, G.~Lugosi, and P.~Massart, \textit{Concentration Inequalities}, Oxford University Press, 2013.

\bibitem{elgamal:book} A.~El Gamal and Y.-H.~Kim, \textit{Network Information Theory}, Cambridge University Press, 2011.

\bibitem{hochwald00:it} B.~M.~Hochwald and T.~L.~Marzetta, ``Unitary space-time modulation for multiple-antenna communications in Rayleigh flat fading,'' \textit{IEEE Trans. Inform. Theory}, 46(2), 543-564, Mar. 2000.

\bibitem{zheng02:it} L.~Zheng and D.~N.~C.~Tse, ``Communication on the Grassmann manifold: a geometric approach to the noncoherent multiple-antenna channel,'' \textit{IEEE Trans. Inform. Theory}, 48(2), 359-383, Feb. 2002.

\bibitem{efron98:book} B.~Efron and R.~Tibshirani, \textit{An Introduction to the Bootstrap}, Chapman \& Hall/CRC, 1998.

\bibitem{gray98:it} R.~M.~Gray and D.~L.~Neuhoff, ``Quantization,'' \textit{IEEE Trans. Inform. Theory}, 44(6), 2325-2383, Oct. 1998.

\end{thebibliography}
\end{document}